\documentclass[aps,prd,twocolumn,showpacs,preprintnumbers,amsmath,amssymb,floatfix,nofootinbib]{revtex4-1}
\usepackage{graphicx}

\usepackage{color}
\usepackage{colordvi}
\usepackage[normalem]{ulem}

\newcommand{\beqn}{\begin{eqnarray}}
\newcommand{\eeqn}{\end{eqnarray}}
\newcommand{\be}{\begin{equation}}
\newcommand{\ee}{\end{equation}}

\newcommand{\bel}[1]{\be\label{#1}}
\newcommand{\ba}{\begin{array}{c}}
\newcommand{\bat}{\begin{array}{cc}}
\newcommand{\ea}{\end{array}}

\newcommand{\bi}{\begin{itemize}}
\newcommand{\ei}{\end{itemize}}

\newcommand{\ket}{\,\rangle}
\newcommand{\bra}{\langle \,}

\newcommand{\Frac}[2]{\frac{\displaystyle #1}{\displaystyle #2}}

\newcommand{\cO}{{\cal O}}

\newcommand{\mF}{\mathcal{F}}
\newcommand{\mG}{\mathcal{G}}
\newcommand{\mH}{\mathcal{H}}

\newcommand{\mO}{\mathcal{O}}

\newcommand{\mT}{\mathcal{T}}

\newcommand{\lsim}{\stackrel{<}{_\sim}}
\newcommand{\gsim}{\stackrel{>}{_\sim}}

\newcommand{\comment}[1]{}

\begin{document}

\title{A bottom-up approach within the electroweak effective theory: \\ constraining heavy resonances }

\author{Antonio Pich${}^{1}$}
\author{Ignasi Rosell${}^{2}$}
\author{Juan Jos\'e Sanz-Cillero${}^{3}$}

\affiliation{${}^1$ IFIC, Universitat de Val\`encia -- CSIC, Apt. Correus 22085, 46071 Val\`encia, Spain }

\affiliation{${}^2$  Departamento de Matem\'aticas, F\'\i sica y Ciencias Tecnol\' ogicas, Universidad Cardenal Herrera-CEU, CEU Universities, 46115 Alfara del Patriarca, Val\`encia, Spain}

\affiliation{${}^3$  Departamento de F\'\i sica Te\'orica and Instituto de F\'\i sica  de  Part\'\i culas  y  del  Cosmos IPARCOS,  Universidad Complutense de Madrid, E-28040 Madrid, Spain}

\begin{abstract}
The LHC has confirmed the existence of a mass gap between the known particles and possible new states. Effective field theory is then the appropriate tool to search for low-energy signals of physics beyond the Standard Model. We adopt the general formalism of the electroweak effective theory, with a non-linear realization of the electroweak symmetry breaking, where the Higgs is a singlet with independent couplings. At higher energies we consider a generic resonance Lagrangian which follows the above-mentioned non-linear realization and couples the light particles to bosonic heavy resonances with $J^P=0^\pm$ and $J^P=1^\pm$. Integrating out the resonances and assuming a proper short-distance behavior, it is possible to determine or to constrain most of the bosonic low-energy constants in terms of resonance masses. Therefore, the current experimental bounds on these bosonic low-energy constants allow us to constrain the resonance masses above the TeV scale, by following a typical bottom-up approach, {\it i.e.}, the fit of the low-energy constants to precise experimental data enables us to learn about the high-energy scales, the underlying theory behind the Standard Model.
\end{abstract}



\maketitle

\vspace{-5cm}

\section{Introduction}

The LHC has confirmed the success of the Standard Model (SM) with the discovery of a Higgs-like\footnote{Although it
might not be the SM Higgs boson, we will refer to this particle as ``Higgs''.} particle~\cite{higgs}, with couplings compatible with the SM expectations, and the non-observation of new states, which establishes the existence of a mass gap between the SM and possible new physics (NP) fields. This gap justifies the use of effective field theories to analyze the data, and the lack of information about the hypothetical underlying theory behind the SM invites us to follow a bottom-up approach, that is, to search for fingerprints of heavy scales at low energies in a systematic way.

In this bottom-up approach the low-energy constants (LECs), or Wilson coefficients, are free parameters which encode the information about the heavy scales, whereas the construction of the effective Lagrangian (the local operators) depends on the light-particle content, the symmetries and the power counting. There is no doubt about the particle content in this case, the SM states, but there are two different ways of introducing the Higgs field, and this has consequences in the symmetries and in the power counting to be used~\cite{Buchalla:2016bse,Pich:2018ltt}. One can consider the more common linear realization of the electroweak symmetry breaking (EWSB), assuming the Higgs to be part of a doublet together with the three electroweak (EW) Goldstones, as in the SM, or the more general non-linear realization, without assuming any specific relation between the Higgs and the Goldstone fields. The first option is known as the SM effective field theory (SMEFT) and it is organized as an expansion in canonical dimensions, being its leading-order (LO) approximation the dimension-four SM Lagrangian. We follow here the second option, the EW effective theory (EWET), also known as Higgs effective field theory (HEFT) or EW chiral Lagrangian (EWChL), where an expansion in generalized momenta is followed. The LO description is given in this case by the Higgsless SM Lagrangian plus the $\cO(p^2)$ operators that introduce the Higgs and the EW Goldstones. Note that the SMEFT is a particular case of the more general EWET framework.

At higher energies we introduce massive resonance states by using a phenomenological Lagrangian respecting the non-linear realization of the EWSB, {\it i.e.}, respecting the symmetries and following the chiral expansion  of the EWET. 

Therefore, we consider two effective Lagrangians with different particle contents: the EWET at low energies, with only the SM particles, and the EW resonance theory at high energies, with the SM particles plus heavy resonances. Both Lagrangians can be matched in a common validity region by integrating out the resonances; in other words, the EWET LECs can be determined in terms of resonance parameters. In order to obtain interesting constraints 
(relevant from a phenomenological point of view), it is very convenient to assume a given short-distance behavior of the unknown underlying theory. This allows us to get determinations or bounds in terms of only resonance masses. 

The main purpose of this work is to combine the current experimental bounds on the bosonic EWET LECs with their determinations or limits in terms of resonance masses, in order to 
constrain the scale of new physics. We update the results already presented in Ref.~\cite{Pich:2015kwa}, extending the resonance Lagrangian (considering $P$-even and also $P$-odd operators, bosonic and also fermionic resonances, color singlets and also color octets), softening the high-energy constraints and including the current experimental bounds.
Moreover, since precise experimental measurements of the $hWW$ coupling $\kappa_W$ are now
available, we no-longer consider this coupling as a non-free parameter.\footnote{In Ref.~\cite{Pich:2015kwa}, $\kappa_W$ was determined theoretically with the second Weinberg Sum Rule~\cite{WSR} of the $W^3B$ correlator at next-to-leading order (NLO): $\kappa_W=M_V^2/M_A^2$~\cite{ST}. This constraint was a consequence of having considered only the two-Goldstone ($\varphi\varphi$) and the Higgs-Goldstone ($h \varphi$) cuts in the NLO determination of the oblique parameters $S$ and $T$~\cite{Peskin_Takeuchi}.} 

Experimental analyses of effective contact interactions in high-energy colliders and direct resonance searches based on Drell-Yan production seem to discard large contributions from four-fermion operators~\cite{lagrangian_color}, not to mention the tight constraints from flavour observables. Hence, the goal of this article is only to test some potentially more sizable bosonic LEC effects. A discussion about current phenomenological limits on four-fermion operators can be found in Ref.~\cite{lagrangian_color}.

The theoretical framework is briefly presented in Section~\ref{sec:lagrangian}: the Lagrangians of the EWET and the resonance electroweak theory are given in Sections~\ref{sec:EWET} and \ref{sec:RET}, respectively, whereas the assumed short-distance behavior is described in Section~\ref{sec:short-distance}. Some technical details are relegated to the Appendix. In Section~\ref{sec:exp} the current experimental bounds on the bosonic EWET LECs are summarized. The connection between the theoretical predictions of Section~\ref{sec:lagrangian} and the experimental bounds of Section~\ref{sec:exp} is done in Section~\ref{sec:pheno}, being the plots in  Figures~\ref{plots1} and ~\ref{plots2} the main results of our analysis. Some concluding remarks are finally given in Section~\ref{sec:conclusions}.

\section{Theoretical Framework}\label{sec:lagrangian}

\subsection{Low energies: EWET Lagrangian} \label{sec:EWET}

\begin{table*}
	\begin{center}
		\renewcommand{\arraystretch}{2.6}
		\begin{tabular}{|c|c|c||c|c|c|}
                  \hline
$i$ & ${\cal O}_i$ & $\mF_i$ & $i$ & ${\cal O}_i$ & $\mF_i$ \\
\hline 
$1$  & $\Frac{1}{4}\,\bra {f}_+^{\mu\nu} {f}_{+\, \mu\nu}- {f}_-^{\mu\nu} {f}_{-\, \mu\nu}\ket_2$ & $- \Frac{F_V^2-\widetilde{F}_V^2}{4M_{V^1_3}^2} + \Frac{F_A^2-\widetilde{F}_A^2}{4M_{A^1_3}^2} $ 
&%
$7$ & 	$\Frac{(\partial_\mu h)(\partial_\nu h)}{v^2} \,\bra u^\mu u^\nu \ket_2$ & $\Frac{ d_P^2}{2 M_{P^1_3}^2}+\Frac{\lambda_1^{hA\,\, 2}v^2}{M_{A^1_3}^2} +  \Frac{\widetilde{\lambda}_1^{hV\,\, 2}v^2}{M_{V^1_3}^2} $ 
\\ [1ex] \hline
$2$  & 	$ \Frac{1}{2} \,\bra {f}_+^{\mu\nu} {f}_{+\, \mu\nu} + {f}_-^{\mu\nu} {f}_{-\, \mu\nu}\ket_2$& $- \Frac{F_V^2+{\widetilde{F}_V}^2}{8M_{V^1_3}^2} - \Frac{F_A^2+{\widetilde{F}_A}^2}{8M_{A^1_3}^2}$ 
&%
$8$ &  $\Frac{(\partial_\mu h)(\partial^\mu h)(\partial_\nu h)(\partial^\nu h)}{v^4}$ & $0$ 
\\ [1ex] \hline
$3$  &	$\Frac{i}{2} \,\bra {f}_+^{\mu\nu} [u_\mu, u_\nu] \ket_2$ & $-  \Frac{F_VG_V}{2M_{V^1_3}^2} - \Frac{\widetilde{F}_A\widetilde{G}_A}{2M_{A^1_3}^2}$ 
&%
$9$ & $\Frac{(\partial_\mu h)}{v}\,\bra f_-^{\mu\nu}u_\nu \ket_2$ & $  - \Frac{F_A \lambda_1^{hA} v}{M_{A^1_3}^2} - \Frac{\widetilde{F}_V \widetilde{\lambda}_1^{hV} v}{M_{V^1_3}^2}$ 
\\ [1ex] \hline
$4$  & 	$\bra u_\mu u_\nu\ket_2 \, \bra u^\mu u^\nu\ket_2 $  & $\Frac{G_V^2}{4M_{V^1_3}^2} + \Frac{{\widetilde{G}_A}^2}{4M_{A^1_3}^2} $
&%
$10$ & $\bra \mT u_\mu\ket_2\, \bra \mT u^\mu\ket_2$ & $-\displaystyle\frac{\widetilde{c}_{\mathcal{T}}^2}{2M_{V^1_1}^2}-\displaystyle\frac{c_{\mathcal{T}}^2}{2M_{A^1_1}^2}$ 
\\ [1ex] \hline
$5$  & $  \bra u_\mu u^\mu\ket_2 \, \bra u_\nu u^\nu\ket_2$ & $\Frac{c_{d}^2}{4M_{S^1_1}^2}-\Frac{G_V^2}{4M_{V^1_3}^2} - \Frac{{\widetilde{G}_A}^2}{4M_{A^1_3}^2}$
&%
$11$ & $ \hat{X}_{\mu\nu} \hat{X}^{\mu\nu}$ & $- \Frac{F_{X}^2}{M_{V^1_1}^2} - \Frac{\widetilde{F}_{X}^2}{M_{A^1_1}^2} $
\\ [1ex] \hline
$6$ & $\Frac{(\partial_\mu h)(\partial^\mu h)}{v^2}\,\bra u_\nu u^\nu \ket_2$ & $ - \Frac{\widetilde{\lambda}_1^{hV\,\, 2}v^2}{M_{V^1_3}^2} - \Frac{\lambda_1^{hA\,\, 2}v^2}{M_{A^1_3}^2}$ 
&%
$12$ & $\bra \hat G_{\mu\nu}\,\hat G^{\mu\nu} \ket_3 $ & $  - \Frac{(C_G)^2}{2 M_{V^8_1}^2} - \Frac{(\widetilde{C}_G)^2}{2 M_{A^8_1}^2}  $ 
\\ [1ex] \hline 
\end{tabular}
	\end{center}
	\caption{\small
		 $P$-even operators (${\cal O}_i$) of the bosonic $\cO(p^4)$ EWET Lagrangian and the contributions to their LECs ($\mF_i$) from heavy scalar, pseudo-scalar, vector, and axial-vector exchanges~\cite{lagrangian,lagrangian_color}.  }
	\label{p_even0}
\end{table*}

\begin{table*}
	\begin{center}
		\renewcommand{\arraystretch}{2.6}
		\begin{tabular}{|c|c|c||c|c|c|}
                  \hline
$i$ & $\widetilde{\cal O}_i$ & $\widetilde{\mF}_i$ & $i$ & $\widetilde{\cal O}_i$ & $\widetilde{\mF}_i$ \\ \hline
$1$ & $\Frac{i}{2} \,\bra {f}_-^{\mu\nu} [u_\mu, u_\nu] \ket_2$  & $- \Frac{\widetilde{F}_VG_V}{2M_{V^1_3}^2} - \Frac{F_A\widetilde{G}_A}{2M_{A^1_3}^2}$ 
& %
$3$ & $\Frac{(\partial_\mu h)}{v}\,\bra f_+^{\mu\nu}u_\nu \ket_2$ & $- \Frac{F_V \widetilde{\lambda}_1^{hV} v}{M_{V^1_3}^2} - \Frac{\widetilde{F}_A \lambda_1^{hA} v}{M_{A^1_3}^2}$ 
\\ [1ex] \hline
$2$ & $\bra {f}_+^{\mu\nu} {f}_{-\, \mu\nu} \ket_2 $  & $- \Frac{F_V \widetilde{F}_V}{4M_{V^1_3}^2} - \Frac{F_A \widetilde{F}_A}{4M_{A^1_3}^2}$ & & &  \\ [1ex] \hline
\end{tabular}
	\end{center}
	\caption{\small
		 $P$-odd operators ($\widetilde{\cal O}_i$) of the bosonic $\cO(p^4)$ EWET Lagrangian and the contributions to their LECs ($\widetilde{\mF}_i$) from heavy scalar, pseudo-scalar, vector, and axial-vector exchanges~\cite{lagrangian,lagrangian_color}.  }
	\label{p_odd0}
\end{table*}

The EWET Lagrangian can be organized as an expansion in powers of generalized momenta~\cite{Pich:2018ltt,lagrangian,lagrangian_color,Buchalla,Weinberg}:
\begin{eqnarray}
\mathcal{L}_{\mathrm{EWET}} &=& \sum_{\hat d\ge 2}\, \mathcal{L}_{\mathrm{EWET}}^{(\hat d)}\,. \label{EWET-Lagrangian0}
\end{eqnarray}
Note that, as it has been stressed previously, the operators are not ordered according to their canonical dimensions and one must use instead the chiral dimension $\hat d$, which reflects their infrared behavior at low momenta~\cite{Weinberg}. Consequently, loops are renormalized order by order in this expansion. 

We collect in the Appendix the definitions of the building blocks, used to construct operators invariant under the electroweak symmetry group $\mG$, and the power-counting rules that determine their chiral dimensions. The relevant bosonic part of the LO EWET Lagrangian is given by\footnote{An alternative notation $a=\kappa_W$, $b=c_{2V}$ is used in some works.}
\begin{eqnarray}
\Delta \mathcal{L}_{\mathrm{EWET}}^{(2)} & =&  \frac{v^2}{4}\,\left( 1 +\Frac{2\,\kappa_W}{v} h + \frac{c_{2V}}{v^2} \,h^2\right) \bra u_\mu u^\mu\ket_2    \, , \label{LOEWET_lagrangian}
\end{eqnarray}
where $h$ denotes the Higgs field, the $u_\mu$ tensor contains the EW Goldstones and $\langle\cdots\rangle_2$ indicates an $SU(2)$ trace. Thus, $\kappa_W$ parametrizes the
$hWW$ coupling in SM units.
Assuming invariance under CP transformations, the bosonic NLO EWET Lagrangian reads~\cite{lagrangian,lagrangian_color}:\footnote{ 
For $h=0$, these $\mF_j$ are related to the $a_i$ couplings of the Higgsless Longhitano Lagrangian~\cite{Herrero:1993nc,Longhitano} in the form $a_i=\mF_i$ for $i=1,4,5$, $a_2=(\mF_3-\widetilde{\mF}_1)/2$ and $a_3=-(\mF_3+\widetilde{\mF}_1)/2$.}
\begin{eqnarray}
\Delta \mathcal{L}_{\mathrm{EWET}}^{(4)} & =&
\sum_{i=1}^{12} \mF_i(h/v)\; \mO_i \, +\, \sum_{i=1}^{3}\widetilde\mF_i(h/v)\; \widetilde \mO_i   \, . \label{EWET_lagrangian}
\end{eqnarray}
Tables~\ref{p_even0} and \ref{p_odd0} display the explicit list of operators 
$\mO_i$ ($P$-even) and $\widetilde \mO_i$ ($P$-odd), respectively.

Since the Higgs is a singlet under $\mG$, the operators of (\ref{EWET_lagrangian}) can be multiplied by arbitrary analytical functions of $h/v$~\cite{Grinstein:2007iv}, {\it i.e.}, their LECs are actually Higgs-dependent functions that can be Taylor-expanded in powers of $h/v$. Given the current experimental status, from now on we consider only the first term in this expansion: $\mF_i\equiv\mF_i (h=0)$ and $\widetilde\mF_i\equiv\widetilde\mF_i(h=0)$. 
The couplings $\mF_2$, $\mF_{11}$, $\mF_{12}$ and $\widetilde{\mF}_2$ are not measurable, because 
their corresponding operators reduce to terms already present in the electroweak and strong Yang-Mills Lagrangians. Therefore, they can be re-absorbed through a renormalization of the gauge couplings $g$, $g'$ and $g_s$. Within the SM, $\kappa_W=c_{2V}=1$ and $\mathcal{F}_{i\neq 2,11,12}=\widetilde\mF_{i\neq 2}=0$.

The operator $\mathcal{O}_1$ contributes to the $W^\pm$ and $Z$ self-energies and can then be accessed through the measurement of the so-called oblique parameters~\cite{Peskin_Takeuchi}. The trilinear and quartic gauge couplings are sensitive to $\mathcal{O}_{1,3}$ and $\mathcal{O}_{1,3-5}$, respectively,
while $\widetilde{\cal O}_{1}$ can modify both. $\mathcal{O}_{10}$ generates custodial-breaking corrections to all Goldstone Green functions. The remaining operators, $\mathcal{O}_{6-9}$ and $\widetilde{\cal O}_3$, contribute to Goldstone vertices involving Higgs bosons.

\subsection{High energies: resonance electroweak theory} \label{sec:RET}

Although the EWET power counting is not directly applicable to the resonance theory, one can construct the Lagrangian in a consistent phenomenological way, {\it \`a la} Weinberg~\cite{Weinberg}, which interpolates between the low-energy and the high-energy regimes: the appropriate low-energy predictions are generated and a given short-distance behavior is imposed~\cite{RChT}. 

Taking into account that here we are interested in the resonance contributions to the bosonic $\cO(p^4)$ EWET LECs, we only need to consider $\cO(p^2)$ operators with up to one bosonic resonance field~\cite{lagrangian_color,lagrangian}. 
Moreover, we can drop the couplings to fermionic resonance fields because their tree-level exchanges do not contribute to the bosonic LECs \cite{lagrangian_color}. 

Following the notation of Ref.~\cite{lagrangian_color}, the dimension of the resonance representation is indicated with upper and lower indices in the scheme $R_{SU(2)}^{SU(3)}$, where $R$ stands for any of the four possible $J^{PC}$ bosonic states with quantum numbers $0^{++}$ (S), $0^{-+}$ (P), $1^{--}$ (V) and $1^{++}$ (A). The normalization used for the n-plets of resonances is
\begin{equation}
R^{n}_3 \,=\, \frac{1}{\sqrt{2}}\,\sum_{i=1}^3\,\sigma_{i}\,R^n_{3,i}\,, \qquad R_n^8 \,=\,  \sum_{a=1}^8\,T^a\, R_n^{8,a}\,,
\end{equation}
with $\bra \sigma_i\sigma_j\ket_2 = 2\delta_{ij}$ and $\bra T^a T^b\ket_3 = \delta^{ab}/2$,
where $\langle\cdots\rangle_3$ indicates an $SU(3)_C$ trace. 

The spin-1 resonances $V$ and $A$ can be described with either a four-vector Proca field $\hat{R}^\mu$ or with an antisymmetric tensor $R^{\mu \nu}$. 
Here we keep both formalisms because, as it was demonstrated in Ref.~\cite{lagrangian}, the sum of tree-level resonance-exchange contributions from the $\cO(p^2)$ resonance Lagrangian with Proca and antisymmetric spin-1 resonances gives the complete (non-redundant and correct) set of predictions for the $\cO(p^4)$ EWET LECs, without any additional contributions from local operators without resonance fields.

The relevant resonance Lagrangian takes the form~\cite{lagrangian_color}:
\begin{widetext}
\begin{align}
\Delta \mathcal{L}_{\mathrm{RT}} &=\,
 \frac{v^2}{4}\!\left(\! 1 \!+\!\Frac{2\,\kappa_W}{v} h \!+\! c_{2V} \,h^2\!\right)\! \bra u_\mu u^\mu\ket_2 
   + \Frac{c_{d}}{\sqrt{2}}\, S^1_1\bra u_\mu u^\mu \ket_2  + d_P \Frac{(\partial_\mu h)}{v} \bra  P^1_3  \, u^\mu \ket_2
 +\widetilde{c}_{\mathcal{T}}\,  \hat V^1_{1\,\mu}  \bra u^\mu \mathcal{T} \ket_2   + c_{\mathcal{T}} \, \hat A^1_{1\,\mu}  \bra u^\mu \mathcal{T} \ket_2  \phantom{\bigg( }
 \nonumber \\
&+\bra V^1_{3\,\mu\nu} \left( \Frac{F_V}{2\sqrt{2}}  f_+^{\mu\nu} + \Frac{i G_V}{2\sqrt{2}} [u^\mu, u^\nu]  + \Frac{\widetilde{F}_V }{2\sqrt{2}} f_-^{\mu\nu}  +  \sqrt{2}\, \widetilde{\lambda}_1^{hV}   (\partial^\mu h) u^\nu  \right) \ket_2  +   F_{X} V^1_{1\,\,\mu\nu}  \hat X^{\mu\nu} + C_G V^8_{1\,\mu\nu}  \hat G^{\mu\nu}\nonumber \\
&+ \bra A^1_{3\,\mu\nu} \left(\Frac{F_A}{2\sqrt{2}}  f_-^{\mu\nu}  + \sqrt{2}\, \lambda_1^{hA}  (\partial^\mu h) u^\nu +  \Frac{\widetilde{F}_A}{2\sqrt{2}} f_+^{\mu\nu} +  \Frac{i \widetilde{G}_A}{2\sqrt{2}} [u^{\mu}, u^{\nu}]    \right) \ket_2  +  \widetilde{F}_{X} A^1_{1\,\mu\nu}  \hat X^{\mu\nu} + \widetilde C_G  A^8_{1\,\mu\nu} \hat G^{\mu\nu} \,.
 \label{Lagrangian}
\end{align} 
\end{widetext}
We only display the terms which contribute to the bosonic LECs we are interested in. 
The first line shows the interactions without resonances, those with a heavy scalar or pseudo-scalar boson, and the heavy vector and axial-vector operators with Proca fields; whereas in the second and third lines the heavy vector and axial-vector contributions with antisymmetric resonances are given, respectively. 

Integrating out the heavy resonance fields in Eq.~(\ref{Lagrangian}), one recovers the EWET Lagrangian (\ref{EWET_lagrangian}) with explicit values of the LECs in terms of resonance parameters. These determinations are also shown in Tables~\ref{p_even0} and \ref{p_odd0}~\cite{lagrangian,lagrangian_color}. 
Note that $\mathcal{F}_4$ and $\mathcal{F}_4+\mathcal{F}_5$ are predicted to be positive, 
in agreement with the known requirements from causality, crossing and analyticity~\cite{Filipuzzi:2012bv}. 

The resonance-exchange contributions to the LECs have the generic structure $\mF_j\sim G_R^2/M_R^2$. The resonance couplings can be easily traded by the corresponding partial decay widths into $\chi\chi=hh,\, \varphi h, \, \varphi\varphi$, which scale as $\Gamma_{R\to\chi\chi}\sim G_R^2 M_R^3/(16\pi v^4) $. The predicted LECs take then the form $\mF_j\sim 16\pi v^4\Gamma_{R\to\chi\chi}/M_R^5$~\cite{Toublan:1995bk,Ecker:2007us,Donoghue:1988ed,Guo:2007ff,Alboteanu:2008my}. Since our low-energy analysis is not able to separate the effects of $\Gamma_{R\to\chi\chi}$ and $M_R$, this expression does not introduce any particular improvement. Nonetheless, writing the observable cross sections in terms of the resonance masses and widths is interesting to orientate present and future collider searches of new-physics states within the few TeV range~\cite{Alboteanu:2008my,Kilian:2015opv,Frank:2012wh,Brass:2018hfw,Delgado:2017cls,Delgado:2019ucx}.

\subsection{Short-distance constraints} \label{sec:short-distance}

As it has been spotlighted previously, assuming a good short-distance behavior is important because of two reasons: 
\begin{itemize}
\item From a theoretical point of view, the resonance theory tries to interpolate between the low-energy and the high-energy regimes: by construction the low-energy effective Lagrangian (the EWET in our case) is recovered when the resonances are integrated out, as it has been shown for the bosonic operators in the previous subsection; however, and in order to deal with a good interpolation with the underlying dynamical theory at high energies, short-distance constraints are needed to implement the assumed high-energy limits. 
\item Moreover, and from a practical point of view, these constraints are very convenient to reduce the number of resonance parameters. In other words, without short-distance constraints Tables~\ref{p_even0} and \ref{p_odd0} show the determination of the fifteen EWET LECs of (\ref{EWET_lagrangian}) in terms of the sixteen resonance couplings of (\ref{Lagrangian}) and the four resonance masses, so the interest of well-motivated constraints is evident.
\end{itemize}

The following high-energy constraints have been considered~\cite{lagrangian}:
\begin{enumerate}
\item Well-behaved form factors. The two-Goldstone and Higgs-Goldstone matrix elements of the vector and axial currents can be characterized by the corresponding vector and axial form factors. Assuming that they vanish at high energies, four constraints are found:
\begin{enumerate}
\item Vector form factor to two EW Goldstones ($\varphi\varphi$):
\begin{equation}
v^2\, -\,F_V\,G_V\,   -\,\widetilde{F}_A\,\widetilde{G}_A \,=\, 0 \,. \label{VFF1}
\end{equation}
\item Axial form factor to two EW Goldstones ($\varphi \varphi$):
\begin{equation}
\widetilde{F}_V\, G_V  \, +\, F_A\,\widetilde{G}_A \,=\, 0 \,. \label{AFF1}
\end{equation}
\item Axial form factor to a Higgs and one EW Goldstone ($h\varphi$):
\begin{equation}
\kappa_W\,v \, -\,F_A\,\lambda_1^{hA}\,   -\,\widetilde{F}_V\,\widetilde{\lambda}_1^{hV} \,=\, \,0\,, \label{AFF2}
\end{equation}
\item Vector form factor to a Higgs and one EW Goldstone ($h\varphi$):
\begin{equation}
\widetilde{F}_A\, \lambda_1^{hA} \, +\, F_V\,\widetilde{\lambda}_1^{hV} \,=\, 0 \,. \label{VFF2}
\end{equation}
\end{enumerate}
\item Weinberg Sum Rules (WSRs). The $W^3B$ correlator is an order parameter of the EWSB. In asymptotically-free gauge theories it vanishes at short distances as $1/s^3$~\cite{Bernard:1975cd}, implying two superconvergent sum rules, the so-called 1st and 2nd WSRs~\cite{WSR}:
\begin{enumerate}
\item 1st WSR (vanishing of the $1/s$ term):
\begin{equation}
F_V^2 +\widetilde{F}_A^2 - F_A^2 - \widetilde{F}_V^2 \,=\, v^2  \,. \label{1WSR}
\end{equation}
\item 2nd WSR (vanishing of the $1/s^2$ term):
\begin{equation}
 F_V^2 M_V^2 + \widetilde{F}_A^2 M_A^2 - F_A^2 M_A^2 - \widetilde{F}_V^2 M_V^2=0 . \label{2WSR}
\end{equation}
\end{enumerate}
While the 1st WSR is expected to be also fulfilled in gauge theories with nontrivial ultraviolet (UV) fixed points, the validity of the 2nd WSR depends on the particular type of UV theory considered~\cite{1stWSR}.
\end{enumerate}

\begin{table}[tb]  
\begin{center}
\renewcommand{\arraystretch}{2.6}
\begin{tabular}{|c|c|c|}
\hline
& \multicolumn{2}{|c|}{$\mF_i$} \\ \hline
$i$ & with 2nd WSR & without 2nd WSR
\\ \hline\hline
$1$ &$-\displaystyle\frac{v^2}{4} \left( \frac{1}{M_V^2} \!+\! \frac{1}{M_A^2} \right)$ & $\!-\displaystyle\frac{v^2}{4M_V^2}\! -\!\frac{F_A^2}{4} \!\left(\!\frac{1}{M_V^2}\!-\!\frac{1}{M_A^2}\!\right) < 
\displaystyle\frac{-v^2}{4M_V^2}$  \\[1ex] \hline%
$3$ & \multicolumn{2}{|c|}{$-\displaystyle\frac{v^2}{2M_V^2}$} 
\\[1ex] \hline
$4$ & $\displaystyle\frac{v^2}{4} \left( \frac{1}{M_V^2}\!-\!\frac{1}{M_A^2}\right)$& $\cdots$ 
\\[1ex] \hline 
$5$ & \multicolumn{2}{|c|}{$\displaystyle\frac{c_d^2}{4M_{S^1_1}^2}-\mathcal{F}_4$}
\\[1ex] \hline
$6$ &  $-\displaystyle \kappa_W^2 v^2\left(\frac{1}{M_V^2}\! -\! \frac{1}{M_A^2} \right)$& 
$\cdots$ 
\\[1ex] \hline
$7$ &  \multicolumn{2}{|c|}{$\displaystyle\frac{d_P^2}{2M_P^2} -\mathcal{F}_6$}
\\[1ex] \hline
 $9$  &   \multicolumn{2}{|c|}{$-\displaystyle\frac{\kappa_W v^2}{M_A^2} $}
\\[1ex] \hline
\end{tabular}
\caption{{\small
 Resonance-exchange contributions to the $P$-even bosonic $\cO(p^4)$ LECs, considering only $P$-even operators and the short distance constraints. Entries marked with $\cdots$ indicate that the result is the same as in Table~\ref{p_even0}, without further simplification.}}
\label{p_even}
\end{center}
\end{table}

\begin{table}[t]  
\begin{center}
\renewcommand{\arraystretch}{2.6}
\begin{tabular}{|c|c|c|}    
\hline
& \multicolumn{2}{|c|}{$\mF_i$} \\ \hline
$i$ & with 2nd WSR & without 2nd WSR
\\ \hline\hline
1 &$\quad  -\displaystyle\frac{v^2}{4} \left( \frac{1}{M_V^2} + \frac{1}{M_A^2} \right)$ & $ -\displaystyle\frac{v^2}{4M_V^2} -\frac{F_A^2-\widetilde F_A^2}{4}  \left( \frac{1}{M_V^2}\! -\!\frac{1}{M_A^2} \right)$
\\ &&
${}^\dagger <  -\displaystyle\frac{v^2}{4M_V^2}$  \\[1ex] \hline
3 &  \multicolumn{2}{|c|}{ $-\displaystyle\frac{v^2}{2M_A^2} - \frac{F_VG_V}{2} \left( \frac{1}{M_V^2} \!-\! \frac{1}{M_A^2} \right)
\quad {}^\ddagger < \quad -\displaystyle\frac{v^2}{2M_A^2}
$}
\\[1ex] \hline
5 &  \multicolumn{2}{|c|}{ $ \displaystyle\frac{c_d^2}{4M_{S^1_1}^2} -\mathcal{F}_4 \phantom{\Bigg)}$}
\\[1ex] \hline
7 &  \multicolumn{2}{|c|}{ $ \displaystyle\frac{d_P^2}{2M_P^2} -\mathcal{F}_6$ }
\\[1ex] \hline
9 &    \multicolumn{2}{|c|}{ $-\displaystyle\frac{\kappa_W v^2}{M_V^2} + F_A \lambda_1^{hA}v \left(  \frac{1}{M_V^2}\!-\!\frac{1}{M_A^2} \right)
\quad {}^\mathsection > \quad -\displaystyle\frac{\kappa_W v^2}{M_V^2}$ }
\\[1ex] \hline
\end{tabular}
\caption{{\small
Resonance-exchange contributions to the $P$-even bosonic $\cO(p^4)$ LECs ($P$-even and $P$-odd operators included), once the short distance constraints are considered. 
The inequalities $^\dagger$, $^\ddagger$ and $^\mathsection$ assume 
that $F_A^2>\widetilde F_A^2$, $F_VG_V>0$ and $F_A \lambda_1^{hA}>0$, respectively.} 
}
\label{p_even_odd1}
\end{center}
\end{table}

Using these constraints, we can sharpen the determinations of the EWET LECs in Tables~\ref{p_even0} and \ref{p_odd0}, writing them in terms of a smaller number of resonance parameters. Combining the two WSRs, one gets the identities
\be\label{eq:FVFA-WSRs}
F_V^2 - \widetilde{F}_V^2\, =\, \frac{v^2 M_A^2}{M_A^2-M_V^2}\, ,
\qquad
F_A^2 - \widetilde{F}_A^2 \,=\, \frac{v^2 M_V^2}{M_A^2-M_V^2}\, .
\ee
In the absence of $P$-odd couplings, these relations fix $F_V$ and $F_A$ in terms of the vector and axial-vector masses and, moreover, imply that $M_A>M_V$. This mass hierarchy remains valid if $\widetilde{F}_V^2 < F_V^2$ and $\widetilde{F}_A^2 < F_A^2$, which is a reasonable working assumption that we will adopt. Notice that Eq.~(\ref{eq:FVFA-WSRs}) implies that $F_V^2 - \widetilde{F}_V^2$ and $F_A^2 - \widetilde{F}_A^2$ must have the same sign. We will also assume that the inequality $M_A>M_V$ is fulfilled in all scenarios, even when the 2nd WSR does not apply.

The predictions obtained when only $P$-even operators are considered\footnote{In this case all terms with tilde in Eqs.(\ref{EWET_lagrangian}), (\ref{Lagrangian})--(\ref{eq:FVFA-WSRs}), (\ref{running}), and in Tables~\ref{p_even0} and \ref{p_odd0} should be discarded.} are shown in Table~\ref{p_even} (most of these results can be found in Ref.~\cite{Pich:2015kwa}). The more general results, including contributions from both $P$-even and $P$-odd operators, are displayed in Table~\ref{p_even_odd1}.
In order to ease the notation, and taking into account that most contributions come from resonances in EW triplets and in QCD singlets ($R^1_3$), neither superindices nor subindices are used in this case in Tables~\ref{p_even} and~\ref{p_even_odd1}, and from now on; that is, $M_R\equiv M_{R^1_3}$, unless something else is specified. 

In Tables~\ref{p_even} and~\ref{p_even_odd1} we only show those determinations of LECs that get improved with the short-distance constraints. 
We consider two types of scenarios: theories where the 2nd WSR applies 
and a more conservative setting where it does not.
The 2nd WSR sharpens considerably the determination of $\mF_1$, and (when only $P$-even contributions are included) also $\mF_4$ and $\mF_6$. In the last two cases, the improvement is lost when the 2nd WSR is dropped, and one gets back to the results in Table~\ref{p_even0}, without further simplification.

In Table~\ref{p_even_odd1} the determinations are necessarily weaker because more unknown couplings are considered. Nevertheless, we can still obtain useful results assuming that the LECs from odd-parity operators are suppressed with respect to those related to even-parity operators.
Assuming $F_A^2>\widetilde F_A^2$ puts an upper bound on $\mF_1$ (indicated with $^\dagger$ in Table~\ref{p_even_odd1}). The assumption $|\widetilde{F}_A\,\widetilde{G}_A|<|F_VG_V|$ implies $F_VG_V  >0$ through Eq.~(\ref{VFF1}), which leads to an upper bound on $\mF_3$ (indicated with $^\ddagger$). Finally, assuming $|\widetilde{F}_V\,\widetilde{\lambda}_1^{hV}| < |F_A\,\lambda_1^{hA}|$ in Eq.~(\ref{AFF2}), one gets $F_A \lambda_1^{hA}>0$, which implies a lower bound on $\mF_9$ (marked with $^\mathsection$).

\section{Experimental constraints} \label{sec:exp}

Here we summarize the strongest experimental constraints on the EWET LECs from current data: 
\begin{enumerate}
\item $\kappa_W$. The ATLAS Collaboration has recently provided the most accurate measurement of the $hWW$ coupling (in SM units),  $\kappa_W=1.05\pm 0.04$~\cite{Aad:2019mbh}, to be compared with the CMS result, $\kappa_W=1.01\pm 0.07$~\cite{Khachatryan:2014jba}. These values assume the absence of NP contributions. The naive average of both results gives $\kappa_W=1.04\pm 0.03$. A global fit to the LHC Run-1 and Run-2 Higgs data, within the context of the EWET, has determined $\kappa_W=1.01\pm0.06$~\cite{deBlas:2018tjm}. We will adopt this last value, which implies the 95\% CL range shown in Table~\ref{exp}.

\item $c_{2V}$. The ATLAS Collaboration has given the first experimental bound 
on the hhWW coupling: $-1.02<c_{2V}<2.71$, at $95\%$ confidence level (CL)~\cite{ATLAS:2019dgh}.

\item $\mathcal{F}_1$. This LEC can be determined through its relation with the oblique parameter $S$~\cite{Peskin_Takeuchi}: $S=-16\pi \mathcal{F}_1$~\cite{ST,Herrero:1993nc,Delgado:2014jda}. 
The Particle Data Group~\cite{Tanabashi:2018oca} quotes the values $S=0.02\pm0.07$ (fixing $U=0$) or $S=0.02\pm 0.10$ (without fixing $U=0$), which translate into 
$-0.003 < \mathcal{F}_1< 0.002$ (fixing $U=0$) or $-0.004 < \mathcal{F}_1 < 0.004$ 
(without fixing $U=0$) at 95\% CL. 
\item $\mathcal{F}_3$. 
The $\gamma W^+W^-$ anomalous triple gauge coupling reads
\begin{equation}
\delta\kappa_\gamma = g^2 \left(
\mathcal{F}_3-\mathcal{F}_1\right)\,=\, \frac{1}{2}\, M_W^2\, 
\frac{f_W+ f_B}{\Lambda^2} \, ,
\end{equation}
where we have also given its expression in terms of SMEFT operators.
Therefore, $\mF_3-\mF_1$ can be directly extracted from the most recent fits of LHC and electroweak precision data, performed within the SMEFT framework \cite{Falkowski:2015jaa,Almeida:2018cld}.
Ref.~\cite{Almeida:2018cld} finds the 95\% CL ranges $-3.0 <f_W/\Lambda^2<3.7$ and $-8.3 < f_B/\Lambda^2 < 26$ (both in units of TeV$^{-2}$).
Taking also into account the previous determination of $\mathcal{F}_1$, one gets $-0.06<\mathcal{F}_3<0.20$ (95\% CL).

\item $\mathcal{F}_4$ and $\mathcal{F}_5$. For these LECs we can use the recent analysis of quartic gauge couplings by the CMS collaboration, in the context of the SMEFT~\cite{Sirunyan:2019der}, which determined $|f_{S0}/\Lambda^4|<2.7\:$TeV$^{-4}$ and $|f_{S1}/\Lambda^4|<3.4\:$TeV$^{-4}$ ($95\%$ CL). Taking into account the relation between these SMEFT LECs and the related EWET LECs~\cite{Rauch:2016pai,claudia,Garcia-Garcia:2019oig},
\begin{equation}
\mathcal{F}_4\,=\, \frac{v^4}{16} \frac{f_{S0}}{\Lambda^4}\,, \qquad \qquad \mathcal{F}_5\,=\, \frac{v^4}{16} \frac{f_{S1}}{\Lambda^4}\,,
\end{equation}
one gets $|\mathcal{F}_4|<0.0006$ and $|\mathcal{F}_5|<0.0008$~\cite{Garcia-Garcia:2019oig},
and combining quadratically both bounds,
$|\mathcal{F}_4+\mathcal{F}_5|<0.0010$. 
These determinations should be taken with some caution because the experimental analysis has neglected potential uncertainties associated with unitarization effects~\cite{Garcia-Garcia:2019oig,Fabbrichesi:2015hsa}.\footnote{A more recent CMS measurement of $WZ$ and $WW$ production in association with two jets finds indeed significantly less stringent constraints when the unitarity condition is taken into account~\cite{Sirunyan:2020gyx}.} We will just consider them here to illustrate how stringent become the bounds on the resonance masses, for that level of precision in $\mF_4$ and $\mF_5$. An improved experimental analysis of these two LECs is needed, given the relevance of unitarity corrections in high-energy vector-boson scattering~\cite{Perez:2018kav,Brass:2018hfw,Delgado:2017cls,Garcia-Garcia:2019oig,Fabbrichesi:2015hsa,Delgado:2019ucx,Szleper:2014xxa} and the possible caveats concerning the validity of the effective theory in collider analyses~\cite{Kalinowski:2018oxd,Kozow:2019txg}.

\end{enumerate}
These results are summarized in Table~\ref{exp}.

\begin{table}[tb]  
\begin{center}
\renewcommand{\arraystretch}{1.2}
\begin{tabular}{|r@{$\,<\,$}c@{$\,<\,$}l|c|c| }
\hline
\multicolumn{3}{|c|}{LEC} & Ref. & Data \\ 
\hline \hline
$0.89$ & $\kappa_W$ & $1.13$  & \cite{deBlas:2018tjm}
& LHC  \\ \hline 
$-1.02$ & $c_{2V}$ & $2.71$ & \cite{ATLAS:2019dgh} & LHC \\ \hline
 $-0.004$  &$ \mathcal{F}_1$& $0.004$  & 
 \cite{Tanabashi:2018oca} & LEP via $S$\\ \hline 
  $-0.06$&$\mathcal{F}_3$ &$0.20$&\cite{Almeida:2018cld} & LEP \& LHC  \\ \hline 
 $-0.0006$&$\mathcal{F}_4$&$0.0006$&\cite{Sirunyan:2019der} & LHC  \\ \hline
$-0.0010$&$\mathcal{F}_4+\mathcal{F}_5$&$0.0010$ &  \cite{Sirunyan:2019der} & LHC  \\  \hline

\end{tabular}
\caption{{\small
Current experimental constraints on bosonic EWET LECs, at 95\% CL.}} \label{exp}
\end{center}
\end{table}

As shown in Tables~\ref{p_even0} and~\ref{p_odd0}, the only contribution from colored resonances to the bosonic $\mO(p^4)$ EWET Lagrangian originates in the exchange of spin-1 color-octet multiplets, $R^8_1$, and goes to the LEC $\mF_{12}(h/v)=\mF_{12}(0)+\frac{h}{v} \mF_{12}'(0)+\mO(h^2)$, multiplying the gluon operator $\bra \hat{G}_{\mu\nu} \hat{G}^{\mu\nu}\ket$. The Higgsless term $\mF_{12}(0)$ is not directly accessible since it just modifies the strong coupling $g_s$. A 95\% CL on the corresponding $hGG$ coupling, $-0.0009<\mF_{12}'(0)<0.0009$, has been extracted from a global EWET fit~\cite{deBlas:2018tjm}. However, we do not consider this bound because, in addition to the $R^8_1$ mass and gluonic coupling, $\mF_{12}'(0)$ is also sensitive to the $h  R^8_1 R^8_1$ and $hg R^8_1 R^8_1$ couplings, preventing any further predictive phenomenology.

\section{Phenomenology} \label{sec:pheno}

\begin{figure*}[!t]
\begin{center}
\begin{minipage}[c]{8.4cm}
\includegraphics[width=8.4cm]{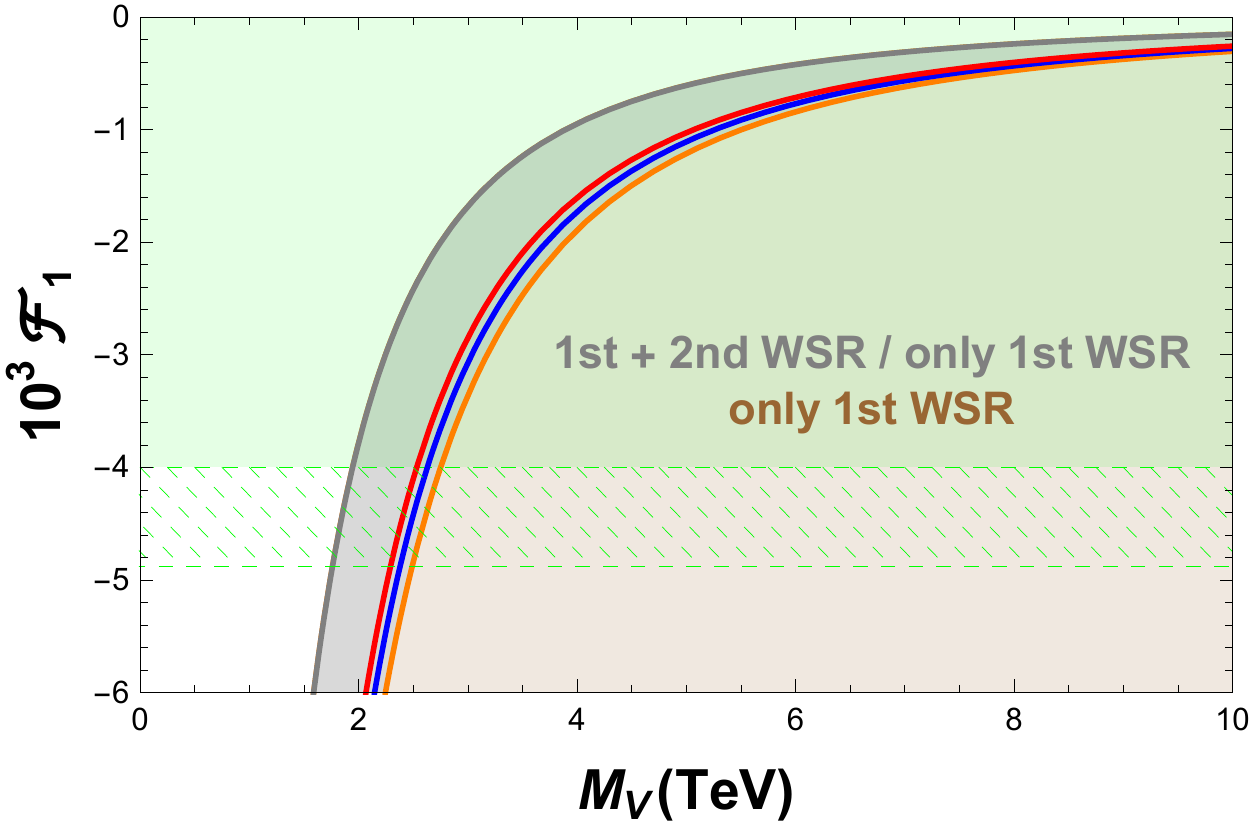}  
\end{minipage}
\hskip .5cm
\begin{minipage}[c]{8.4cm}
\includegraphics[width=8.4cm]{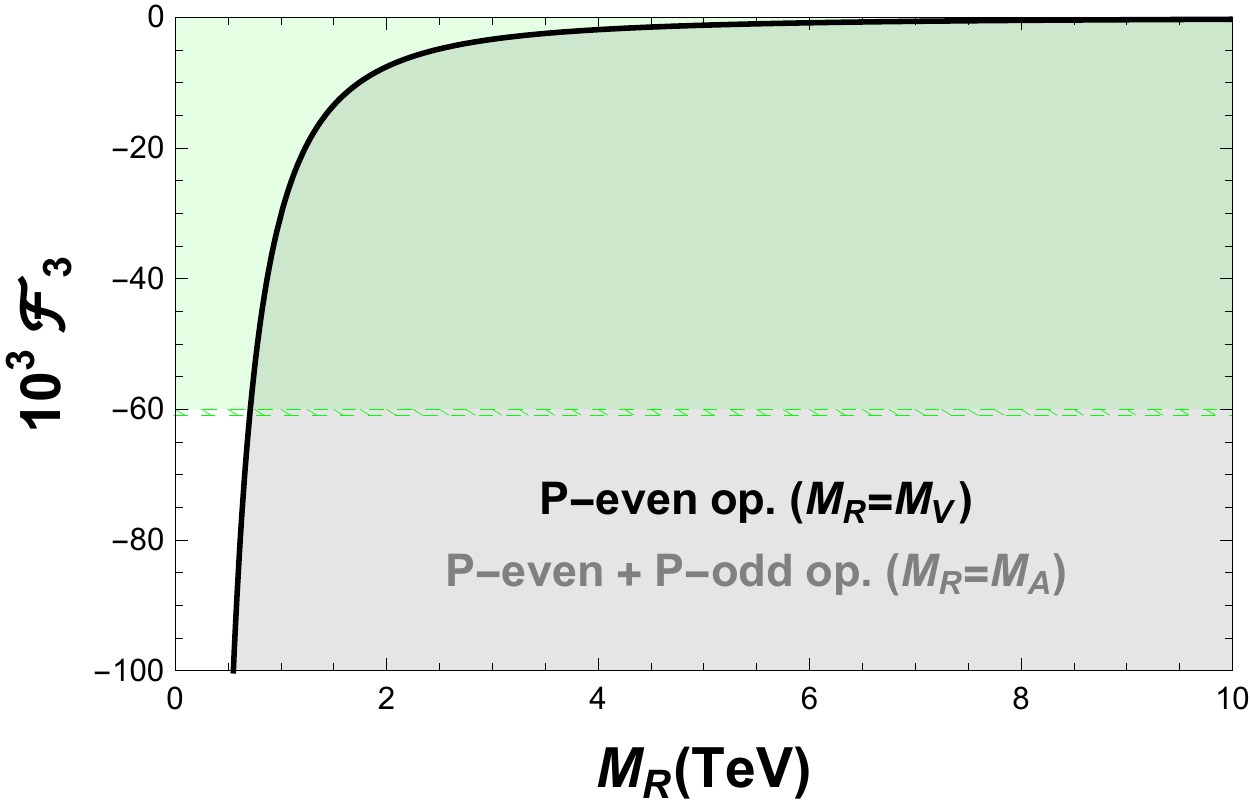} 
\end{minipage}
\\[8pt]
\begin{minipage}[c]{8.4cm}
\includegraphics[width=8.4cm]{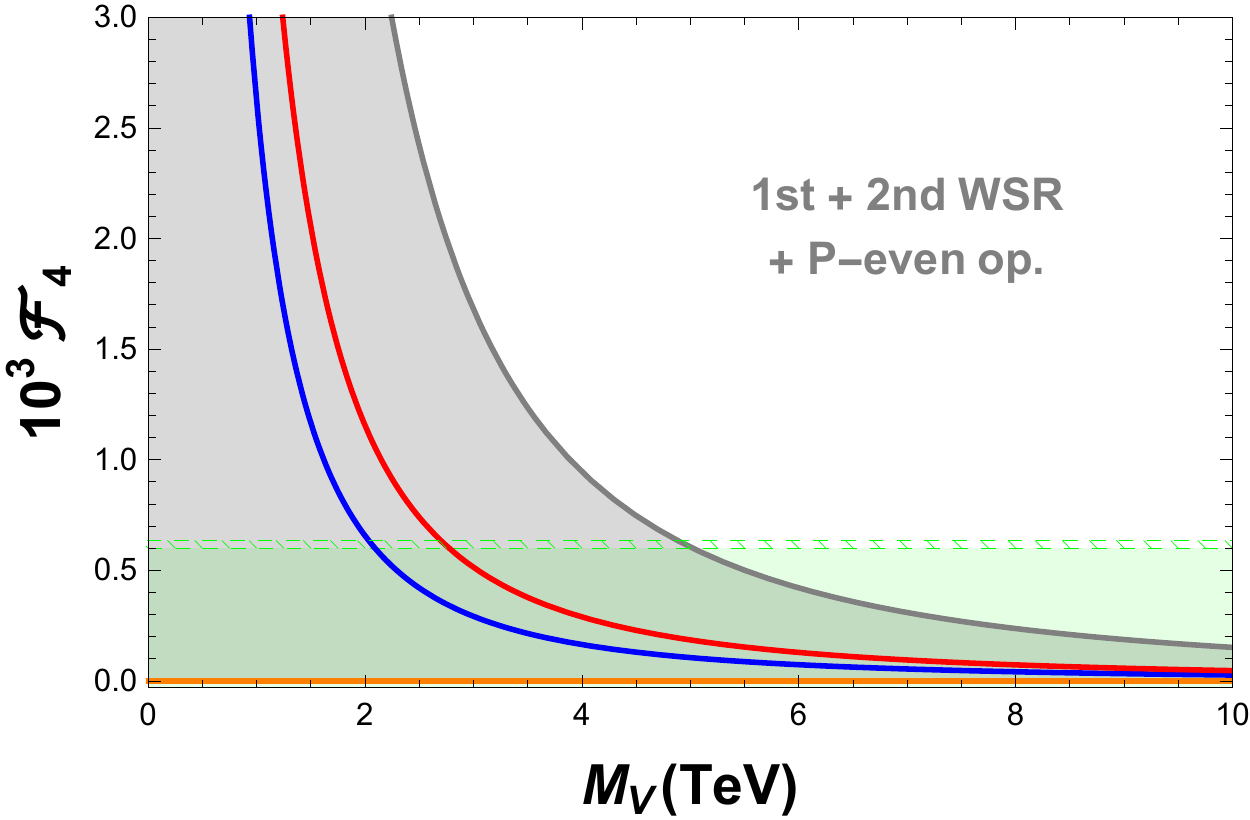} 
\end{minipage}
\hskip .2cm
\begin{minipage}[c]{8.4cm}
\includegraphics[width=8.4cm]{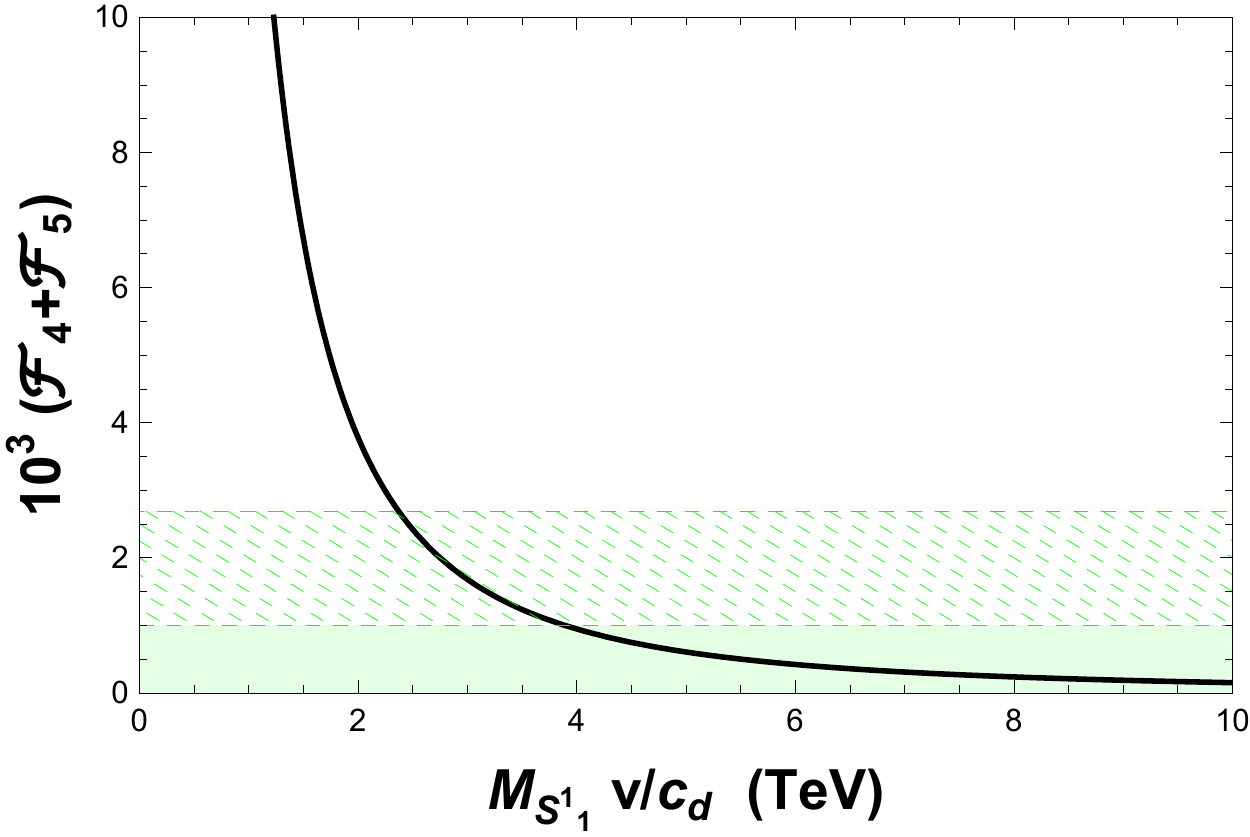}
\end{minipage}
\end{center}
\caption{{\small Predicted values for the LECs $\mF_1$, $\mF_3$, $\mF_4$ and $\mF_4+\mF_5$, 
from Tables~\ref{p_even} and~\ref{p_even_odd1}, 
as a function of the corresponding resonance mass ($M_V$, $M_A$ or $M_{S_1^1} \,v/c_d$).   
The green area covers the experimentally allowed region, at 95\% CL, and it is further extended by a dashed green band that accounts for our estimated one-loop running uncertainties in Eq.~(\ref{eq:1loop-error}). If there is a dependence on $M_V$ and $M_A$, 
the gray and/or brown regions cover all possible values for $M_A>M_V$. If the 2nd WSR has been considered, it is explicitly indicated in the plot, with the corresponding lines for $M_A=M_V$ (orange), $M_A=1.1\,M_V$ (blue), $M_A=1.2\,M_V$ (red) and $M_A\to \infty$ (dark gray). 
In the case without the 2nd WSR, the theoretically allowed region for $\mF_1$ is given by the gray and brown regions. In case of using only the even-parity operators, we indicate it in the plot.
}} 
\label{plots1}
\end{figure*}


\begin{figure*}[!t]
\begin{center}
\begin{minipage}[c]{7.4cm}
\includegraphics[width=7.4cm]{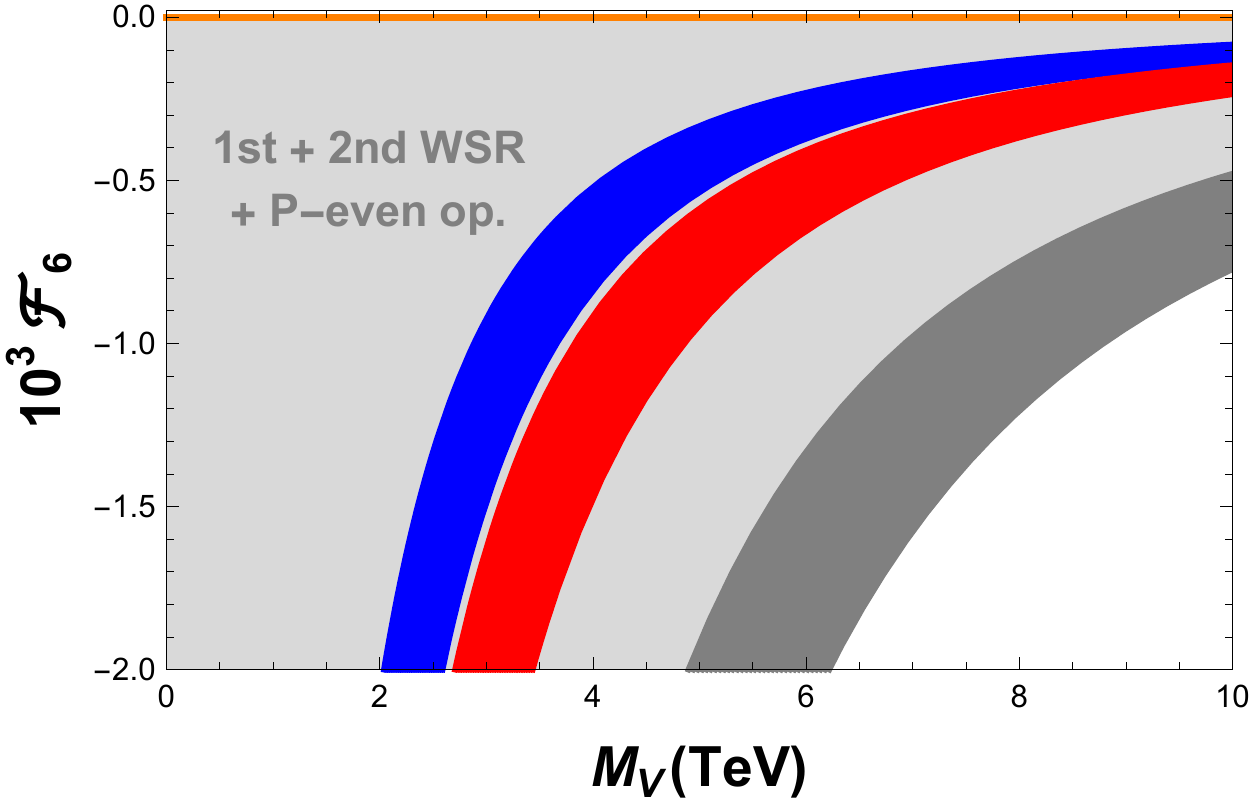} 
\end{minipage}
\begin{minipage}[c]{7.4cm}
\includegraphics[width=7.4cm]{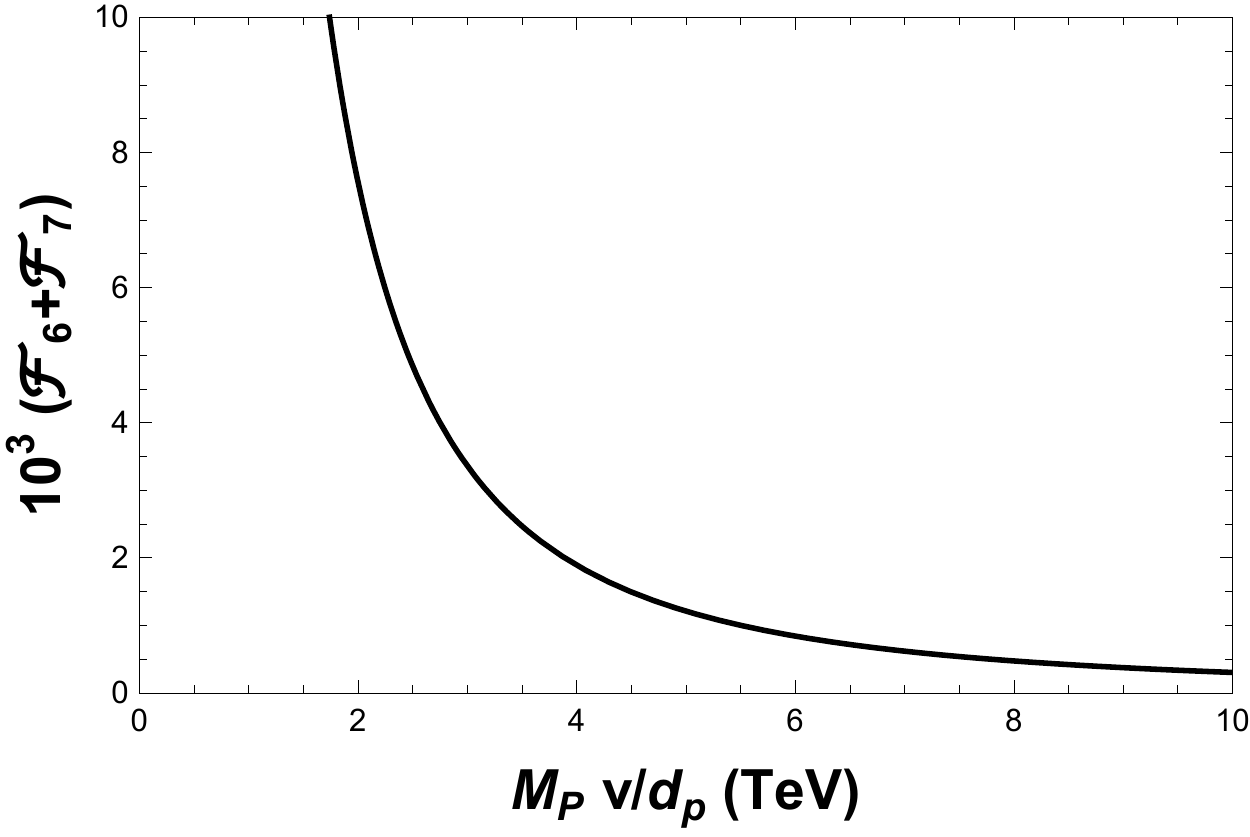}
\end{minipage} 
\\[8pt]\begin{minipage}[c]{7.4cm}
\includegraphics[width=7.4cm]{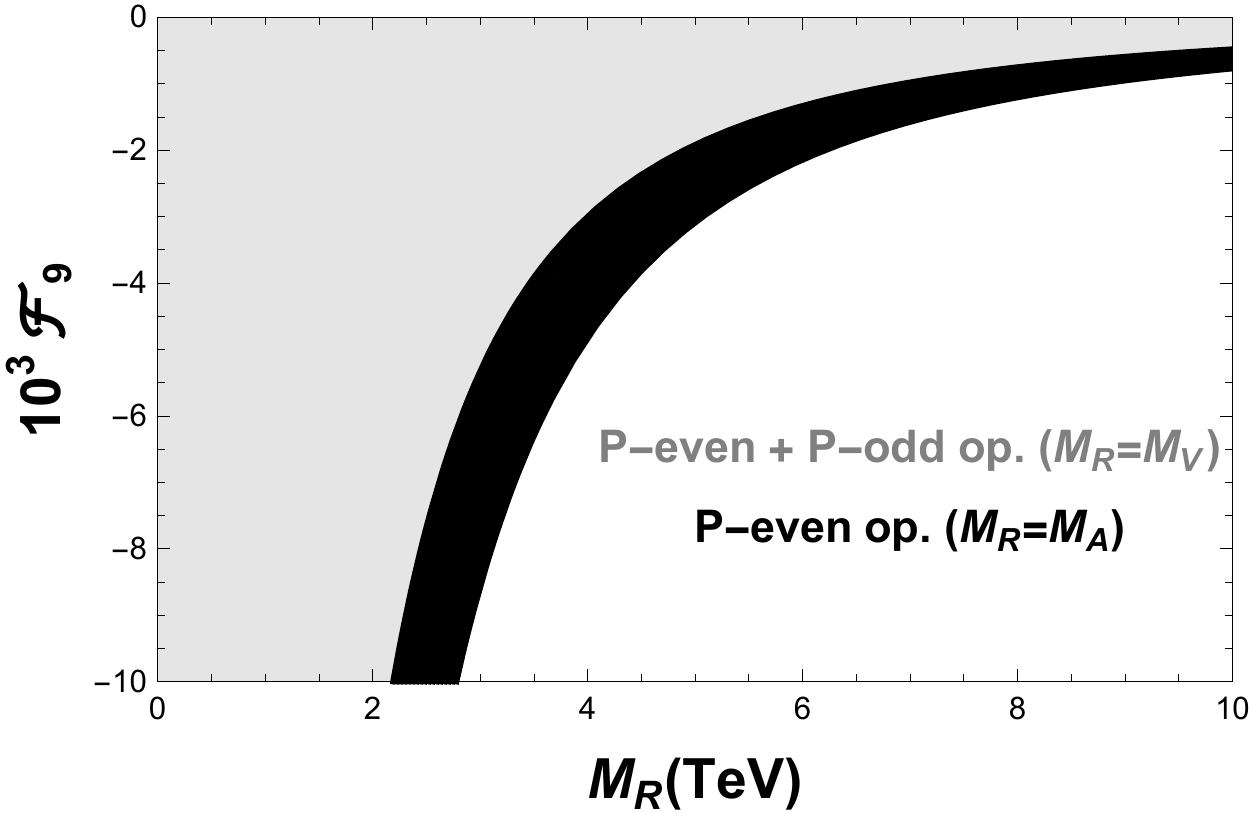} 
\end{minipage}
\end{center}
\caption{{\small Predicted values for the LECs $\mF_6$, $\mF_6+\mF_7$ and $\mF_9$, from Tables~\ref{p_even} and~\ref{p_even_odd1}, as a function of the corresponding resonance mass ($M_V$, $M_A$ or $M_P \,v/d_P$). If there is a dependence on $M_V$ and $M_A$, the gray regions cover all possible values for $M_A>M_V$. 
The $\mF_6$ plot assumes the 2nd WSR and shows the corresponding lines for $M_A=M_V$ (orange), $M_A=1.1\,M_V$ (blue), $M_A=1.2\,M_V$ (red) and $M_A \to \infty$ (dark gray). 
The lines are thick due to the experimental uncertainty on $\kappa_W$. In case of using only the even-parity operators, we indicate it in the plot.
}} \label{plots2}
\end{figure*}


The implementation of short-distance constraints has allowed us to determine some bosonic LECs in terms of very few resonance parameters, as shown in Tables~\ref{p_even} and~\ref{p_even_odd1}. These predictions are plotted in Figures~\ref{plots1} and~\ref{plots2}, as functions of the relevant heavy resonance masses. The green bands in Figure~\ref{plots1} indicate the regions allowed  by the experimental constraints in Table~\ref{exp}. There is still no experimental information available on the LECs plotted in Figure~\ref{plots2}.

The top-left panel in Figure~\ref{plots1} displays the dependence of $\mF_1$ on $M_V$. 
For theories where the 1st WSR is obeyed, the
dark gray curve indicates the predicted upper bound $\mF_1 < -v^2/(4 M_V^2)$. Thus, the whole region below this line (gray and brown areas) would be theoretically allowed  
if only the 1st WSR is assumed. 
In those scenarios where the 2nd WSR is valid, $\mF_1$ is predicted to be a function of $M_V$ and $M_A$, the dark gray curve corresponding to the limit $M_A\to\infty$. The red ($M_A=1.2\, M_V$), blue ($M_A=1.1\, M_V$) and orange ($M_A=M_V$) curves show the predicted values of $\mF_1$ for some representative axial-vector masses, the orange line being the lower bound $\mF_1=-v^2/(2M_V^2)$.
This range of $M_A\sim M_V$ corresponds actually to the most plausible scenario~\cite{Pich:2015kwa}, since the one-loop analysis of the oblique $S$ parameter~\cite{ST} indicates (under very soft and reasonable assumptions) that $\kappa_W\approx M_V^2/M_A^2$ when the 2nd WSR is valid, and the LHC experiments have found indeed that $\kappa_W\approx 1$.

The WSRs do not play any role in $\mF_3$. If one only considers $P$-even operators, $\mF_3 = -v^2/(2 M_V^2)$. This theoretical prediction is shown by the black curve in the top-right panel of Figure~\ref{plots1}. Adding possible $P$-odd contributions, we can only put the upper bound 
$\mF_3 < -v^2/(2 M_A^2)$, which is represented by the same curve but this time with $M_R=M_A$. Thus, the whole region below this line (gray area) would be allowed in the most general case.

The bottom-left panel in Figure~\ref{plots1} shows the predicted values of $\mF_4$, assuming the two WSRs and considering only $P$-even operators, as function of $M_V$. It depends on both $M_V$ and $M_A$, with the upper bound (dark gray curve) obtained at $M_A\to\infty$. Thus, the theoretically allowed region is the gray area below that curve.
The red ($M_A=1.2\, M_V$), blue ($M_A=1.1\, M_V$) and orange ($M_A=M_V$) curves show again the predicted values for different axial masses. The vector and axial-vector contributions have different signs and exactly cancel each other in the equal-mass limit.

Independently of any assumptions concerning WSRs or $P$-odd operators, the contributions from vector and axial-vector resonance exchanges cancel exactly in the combination $\mF_4 + \mF_5 = c_d^2/(4 M_{S^1_1}^2) \equiv v^2/(4 {\hat M}_S^2)$. Thus, one gets a clean prediction that is shown by the black curve in the bottom-right panel of Figure~\ref{plots1},
as function of $\hat M_S = M_{S^1_1} v/c_d$.
A similar cancellation takes place in the combination $\mF_6 + \mF_7 = d_P^2/(2 M_{P}^2) \equiv v^2/(4 {\hat M}_P^2)$, which is plotted in the top-right panel of Figure~\ref{plots2}, as 
function of $\hat M_P = M_{P} v/d_P$.

Although our analysis did not include contributions from spin-2 tensor resonances, their impact on the Higgsless bosonic operators can be easily obtained from previous studies within Chiral Perturbation Theory. Taking into account the short-distance constraints on forward Goldstone-Goldstone scattering, the only LEC sensitive to tensor-exchange is $\mF_5$, which receives a positive contribution~\cite{Toublan:1995bk,Ecker:2007us}.\footnote{Spin-2 tensor studies that do not incorporate short-distance relations lead to tensor contributions to both $\mF_4$ and $\mF_5$, satisfying $\Delta\mF_4^T\geq 0$ and $\Delta\mF_4^T+\Delta \mF_5^T\geq 0$~\cite{Donoghue:1988ed,Alboteanu:2008my}.} Thus, the presence of an exotic tensor resonance would reinforce the positive prediction for $\mF_4 + \mF_5$, shown in Figure~\ref{plots1}. In addition, there are also potential contributions to the $\mF_{6,7,8}$ LECs, which should be explored in future works.  The impact of tensor resonances in LHC searches has been studied in Refs.~\cite{Alboteanu:2008my,Kilian:2015opv,Frank:2012wh}. 

The top-left panel in Figure~\ref{plots2} displays the coupling $\mF_6$, as function of $M_V$, assuming the two WSRs to be valid and considering only $P$-even operators. The theoretical prediction is always negative and depends on $M_V$, $M_A$ and $\kappa_W$, with its lower bound attained at $M_A\to\infty$ (dark gray band). The light-gray area above this bound represents the whole allowed region with $M_A\ge M_V$. The red ($M_A=1.2\, M_V$), blue ($M_A=1.1\, M_V$) and orange ($M_A=M_V$) bands correspond to different axial masses. The thickness of the bands reflects the current experimental 95\% CL uncertainty on $\kappa_W$. The upper bound on $\mF_6$ is obtained at $M_A=M_V$ where the vector and axial-vector contributions exactly cancel.

Finally, the predicted values of $\mF_9$ are shown in the bottom panel in Figure~\ref{plots2}. Considering only $P$-even operators, $\mF_9 = -\kappa_W v^2/M_A^2$, which corresponds to the black band. Its thickness reflects again the experimental error on $\kappa_W$. When $P$-odd operators are taken into account, this prediction transforms into a lower bound that is represented by the same band, but this time as function of $M_R=M_V$. The whole gray area above the band is then theoretically allowed.

Our tree-level predictions from resonance exchange are actually expected to apply at a high scale around the resonance masses, while the experimental constraints on the LECs in Table~\ref{exp} have been obtained at lower energy scales. The one-loop running of the LECs with the renormalization scale is known~\cite{Guo:2015isa}, and the explicit expressions are given in the Appendix, in Eq.~(\ref{running}). These running contributions are of order $1/(4\pi^2) \sim 10^{-3}$ and depend on the LO couplings $\kappa_W$ and $c_{2V}$. They vanish in the SM limit, $\kappa_W=c_{2V}=1$, as they should.
Therefore, for resonances in the few TeV range, $M_R\sim 4\pi v\approx 3$~TeV, we can estimate the potential numerical size of this running effect through the differences
\be 
\Delta \mathcal{F}_i \, =\, |\mathcal{F}_i(\mu\!=\!m_h) - \mathcal{F}_i(\mu\!=\!3\mbox{ TeV})|\, .
\ee
Taking into account the current experimental errors on $\kappa_W$ and $c_{2V}$, we obtain
\begin{eqnarray}
&&\Delta \mathcal{F}_1 = \Delta \mF_3 =  0.9 \cdot 10^{-3}\, ,
\quad\;\;\: 
\Delta \mathcal{F}_{4} =  3 \cdot 10^{-5}\, ,
\nonumber\\
&&\Delta( \mathcal{F}_{4}+\mathcal{F}_{5}) = 1.7 \cdot  10^{-3}\, , 
\qquad\;\;\,    
\Delta \mathcal{F}_{6} =  3\cdot  10^{-3}\, ,
\nonumber\\
&&\Delta (\mathcal{F}_{6}+\mF_7) =  0.6\cdot  10^{-2}\, , 
\qquad  
\Delta \mathcal{F}_{9} =  1.4\cdot  10^{-2}\, . \quad
\label{eq:1loop-error}
\end{eqnarray}
The impact of these running contributions is indicated in Figure~\ref{plots1} with the dashed green bands that enlarge the experimentally allowed regions. Our estimates in Eq.~(\ref{eq:1loop-error}) show that the numerical size of the running uncertainty is much more important for the couplings in Figure~\ref{plots2}, which are not yet constrained experimentally. This is mainly due to the current poor knowledge on the $hhW^+W^-$ coupling, since the anomalous dimensions of these LECs are sensitive to $c_{2V}$: the uncertainties $\Delta \mathcal{F}_{6}$, $\Delta \mathcal{F}_{6+7}$ and $\Delta \mathcal{F}_{9}$  in~(\ref{eq:1loop-error}) strongly decrease, becoming $\cO(10^{-3})$, if we neglect the uncertainty on $ c_{2V}$.  Thus, this error could be sizeably reduced with future data.

\section{Conclusions} \label{sec:conclusions}

Taking into account the great experimental success of the SM, at the currently explored energies, and the emerging evidence about the existence of a mass gap between the SM particles and hypothetical NP states, we have considered a model-independent effective field theory approach to catch any possible deviations from the SM predictions at low energies. Specifically, we have adopted the general non-linear electroweak effective theory (EWET) formalism.

The main aim of this work has been to constrain the scale of the heavy states, which are not directly accessible at current experiments. We have followed a bottom-up approach, where the experimental determination of the LECs of the EWET is used to get imprints of the heavy resonances.

As a consequence, the lightest resonances need to be incorporated in the EWET formalism at higher energies: we have considered here a phenomenological Lagrangian which interpolates  between the low-energy and the high-energy regimes~\cite{lagrangian_color,lagrangian}. In this way, and after integrating out the resonances, the bosonic LECs of the EWET have been  determined in terms of resonance parameters, as it is shown in Tables~\ref{p_even0} and \ref{p_odd0}~\cite{lagrangian_color,lagrangian}. 
These theoretical predictions can be considerably improved by assuming a proper short-distance behavior of the UV theory, which has allowed us to determine or constrain the bosonic LECs in terms of only resonance masses and the $hWW$ coupling $\kappa_W$, as it is shown in Tables~\ref{p_even}  and \ref{p_even_odd1}. 

Combining our theoretical predictions with the current experimental bounds on the bosonic LECs, we have obtained the results shown in Figures~\ref{plots1} and~\ref{plots2}.  
These plots push the resonance mass scale to the TeV range, $M_R \geq 2\,$TeV, in good agreement 
with our previous theoretical estimates in Refs.~\cite{Pich:2015kwa,ST}, based on a NLO calculation of the $S$ and $T$ oblique parameters within a simplified version of the resonance Lagrangian of Eq.~(\ref{Lagrangian}).

The oblique $S$-parameter produces the most precise LEC determination at NLO ($\mF_1$), which implies the resonance-mass lower bounds $M_{V,A}\gsim 2$~TeV, at the 95\% CL.
On the other hand, the anomalous triple gauge couplings provide a much weaker limit on $\mF_3$, which translates in the softer constraint $M_{V,A}\gsim 0.5$~TeV.  

A recent CMS study has led to very stringent bounds on the couplings that rule WW, WZ and ZZ scattering~\cite{Sirunyan:2019der}, 
$|\mF_{4,5}|\lsim 10^{-3}$ in the context of the EWET. In spite of its possible issues regarding unitarity~\cite{Garcia-Garcia:2019oig,Fabbrichesi:2015hsa}, it is illustrative to study the implications of such  level of precision on the anomalous quartic gauge couplings. The limit on $\mF_4+\mF_5$ implies that the singlet scalar resonance would have a mass $M_{S_1^1}\gsim 2$~TeV for a  $S_1^1 WW$ coupling close to the $hWW$ one ($c_d\sim v$). This lower bound would increase if there were additional contributions from spin--2 tensor resonances, since $\Delta \mF_4^T+\Delta \mF_5^T\geq 0$~\cite{Toublan:1995bk,Ecker:2007us,Donoghue:1988ed,Alboteanu:2008my}. Likewise, in the case of BSM extensions with only P-even operators and obeying the two WSRs, the  bounds on $\mF_4$ 
constrain the mass of the vector resonance to the range $M_V\gsim 2$~TeV if $M_A/M_V>1.1$. Nonetheless, lower vector masses would be still allowed if the vector and axial-vector states happened to be very degenerate ($1<M_A/M_V<1.1$).

Currently, there is no data on the remaining NLO LECs. Triplet pseudo-scalar resonances with masses $\hat M_P = M_P v/d_P\lsim 2$~TeV would imply a lower bound $\mF_6+ \mF_7\gsim 5\cdot 10^{-3}$, 
a LEC combination related with $WW\to hh$ at NLO. 
Likewise, a triplet vector resonance with mass
$M_V\sim 2$~TeV leads to the constraint $\mF_6<-2\cdot 10^{-3}$ (also relevant for $WW\to hh$ scattering) for $M_A/M_V>1.1$, in P-even theories with two WSRs. 
Finally, the coupling $\mF_9$, related to the $hWW$ vertex at NLO, could be $\cO(10^{-2})$ in absolute value for $M_{V,A}\sim 2\,$TeV (notice the negative sign in Fig.~\ref{plots2}). However, one-loop corrections introduce corrections of a similar size in $\mF_6$, $\mF_6+\mF_7$ and $\mF_9$, respectively, due to the poor knowledge on the $hhWW$ LO coupling $c_{2V}$. Any further progress on these three LECs requires a similar improvement in the $c_{2V}$ precision and the incorporation of one-loop EWET contributions in these experimental analyses.

In summary, the experimental LHC constraints start already to be competitive. This type of analysis will generate much more precise information, once the expected high-statistics data samples from the upgraded LHC runs will be available.

\vspace{0.3cm}

\acknowledgments

{\bf Acknowledgments}: 
This work has been supported in part by the Spanish Government and ERDF funds from the European Commission (FPA2016-75654-C2-1-P, FPA2017-84445-P, PID2019-108655GB-I00); by the Generalitat Valenciana (PROMETEO/2017/053); by the Universidad Cardenal Herrera-CEU (INDI18/11 and INDI19/15); and by the STSM Grant from COST Action CA16108.

\appendix

\section{Constructing the EWET}
\label{appendix}

At low energies we consider the non-linear EWET Lagrangian, where one has the particle content of the SM, but with the Higgs $h$ as a scalar singlet. The main assumption is that the underlying high-energy theory and the EWET possess the EWSB pattern of the SM:
\begin{equation}
\mG\equiv SU(2)_L\otimes SU(2)_R\quad \longrightarrow \quad \mH\equiv SU(2)_{L+R}\, .
\end{equation}
The remaining symmetry group $\mH$ is the so-called ``custodial'' symmetry \cite{Sikivie:1980hm}, since it protects the ratio of the $W$ and $Z$ masses from large corrections.

The EW Goldstones can be described in the CCWZ formalism~\cite{CCWZ} through the $\mG/ \mH$ coset representative $u(\varphi)=\exp\{ i\vec{\sigma}\,\vec{\varphi}/(2v)\}$, which transforms under the symmetry group element 
$g\equiv (g_L^{\phantom{\dagger}},g_R^{\phantom{\dagger}})\in \mG$ as
\begin{eqnarray}
u(\varphi)\,\rightarrow\, g_L^{\phantom{\dagger}}\, u(\varphi)\,  g_h^\dagger \, =\, g_h^{\phantom{\dagger}} \, u(\varphi) \, g_R^\dagger\, , 
\nonumber\\
U(\varphi)\,\equiv\,  u(\varphi)^2 \,\rightarrow\, g_L^{\phantom{\dagger}}\, U(\varphi)\, g_R^\dagger \, ,\quad
\end{eqnarray}
being the compensating transformation $ g_h^{\phantom{\dagger}}\equiv g_h^{\phantom{\dagger}}(\varphi,g) \in \mH$. 
By promoting $\mG$ to a local symmetry, the auxiliary $SU(2)_L$ and $SU(2)_R$ matrix fields, $\hat{W}_\mu$ and $\hat{B}_\mu$ respectively, and their field-strength tensors are introduced:
\begin{align}
\hat{W}^\mu &\rightarrow\, g_L^{\phantom{\dagger}}\, \hat{W}^\mu g_L^\dagger + i\, g_L^{\phantom{\dagger}}\, \partial^\mu g_L^\dagger\, , \nonumber \\
\hat{B}^\mu &\rightarrow\, g_R^{\phantom{\dagger}}\, \hat{B}^\mu g_R^\dagger + i\, g_R^{\phantom{\dagger}}\, \partial^\mu g_R^\dagger\, , \nonumber \\
\hat{W}_{\mu\nu}  &= \partial_\mu \hat{W}_\nu - \partial_\nu \hat{W}_\mu - i\, [\hat{W}_\mu,\hat{W}_\nu] \, \rightarrow\, g_L^{\phantom{\dagger}}\, \hat{W}_{\mu\nu} \, g_L^\dagger\, , \nonumber \\
\hat{B}_{\mu\nu}  &= \partial_\mu \hat{B}_\nu - \partial_\nu \hat{B}_\mu - i\, [\hat{B}_\mu,\hat{B}_\nu] \,\rightarrow\, g_R^{\phantom{\dagger}}\, \hat{B}_{\mu\nu} \, g_R^\dagger \, , \nonumber \\
f_\pm^{\mu\nu} & =\, u^\dagger \hat{W}^{\mu\nu}  u \pm u\, \hat{B}^{\mu\nu} u^\dagger \,\rightarrow\, g_h^{\phantom{\dagger}}\, f_\pm^{\mu\nu}\, g_h^\dagger \, .
\end{align}
The covariant derivatives are provided by these 
fields:
\begin{align}
&\hskip -.2cm
D_\mu U \, =\, \partial_\mu U -  i\, \hat{W}_\mu  U + i\, U \hat{B}_\mu \,\rightarrow\, g_L^{\phantom{\dagger}}\, (D_\mu U) \, g_R^\dagger \, , \nonumber \\
u_\mu&  =  i\, u\, (D_\mu U)^\dagger u  = -i\, u^\dagger D_\mu U\, u^\dagger  = u_\mu^\dagger \,\rightarrow\, g_h^{\phantom{\dagger}}\, u_\mu\, g_h^\dagger  , 
\end{align}
The identification \cite{Pich:2012jv}
\bel{eq:SMgauge}
\hat{W}^\mu \, =\, -g\;\frac{\vec{\sigma}}{2}\, \vec{W}^\mu \, ,
\qquad\qquad
\hat{B}^\mu\, =\, -g'\;\frac{\sigma_3}{2}\, B^\mu\, ,
\ee
explicitly breaks the chiral symmetry group $\mG$ while preserving the $SU(2)_L\otimes U(1)_Y$ gauge symmetry, as in the SM.

Once the fermion doublets are considered, the fields $\hat{G}_\mu$ and $\hat{X}_\mu$ are introduced to keep the covariance under local $SU(3)_C$ and $U(1)_X$ transformations, respectively. The definitions of these fields, their field-strength tensors $\hat G_{\mu\nu}$ and $\hat{X}_{\mu\nu}$ and the identifications required to break $\mG$ while preserving the $SU(3)_C\otimes SU(2)_L\otimes U(1)_Y$ gauge symmetry can be found in Ref.~\cite{lagrangian_color}.

The explicit breaking of custodial symmetry can be incorporated  by means of a right-handed spurion:
\bel{eq:T_covariant}
\mT_R\,\rightarrow\, g_R^{\phantom{\dagger}}\, \mT_R\, g_R^\dagger\, ,
\qquad  \mT\, =\, u\, \mT_R\, u^\dagger \,\rightarrow\, g_h^{\phantom{\dagger}} \mT g_h^\dagger\, .
\ee
The identification
\begin{equation}
\label{eq:T_R-value}
   \mT_R    \, =\, -g'\;\frac{\sigma_3}{2}\, ,
\end{equation}
allows one to obtain the custodial symmetry breaking operators induced through quantum loops with internal $B_\mu$ lines. 

The power counting of chiral dimensions adopted to organize the operators of the EWET can be summarized as: $h  \sim  \cO\left(p^0\right)$, $ u_\mu, \partial_\mu , \mT \sim  \cO\left( p^1\right)$ and $f_{\pm\, \mu\nu}, \hat{G}_{\mu\nu}, \hat{X}_{\mu\nu} \sim  \cO\left( p^2\right)$~\cite{lagrangian,lagrangian_color}.

The $\mO(p^4)$ operators in Eq.~(\ref{EWET_lagrangian}) renormalize the UV divergences from one-loop diagrams with LO vertices. The running of the renormalized parameters $\mathcal{F}_i$ and $\widetilde\mF_i$,
\begin{equation}
\frac{\partial \mathcal{F}_i}{\partial \ln \mu} = - \frac{\Gamma_i}{16\pi^2}\, , 
\qquad\qquad
\frac{\partial \widetilde\mF_i}{\partial \ln \mu} = - \frac{\widetilde{\Gamma}_i}{16\pi^2} \,,
\label{running}
\end{equation}
has been calculated using the background field method~\cite{Guo:2015isa}:
\begin{align}
\Gamma_1 &=\Gamma_3 =-\frac{1}{6} \left(1\!-\!\kappa_W^2 \right) ,
\qquad
\Gamma_2 = -\frac{1}{12} \left(1\!+\!\kappa_W^2 \right) , 
\nonumber \\
\Gamma_4 &=\frac{1}{6} \left(1\!-\!\kappa_W^2 \right)^2\!,\quad
\Gamma_5 =\frac{1}{8}\left(\kappa_W^2\!-\!c_{2V}\right)^2\!+\!\frac{1}{12} \left(1\!-\!\kappa_W^2 \right)^2 \!,
\nonumber  \\
\Gamma_6 &=-\frac{1}{6} \left(\kappa_W^2\!-\!c_{2V}\right) \left( 7\kappa_W^2\!-\!c_{2V}-6 \right)  , 
\nonumber \\
\Gamma_7 &= \frac{4}{9}\, \Gamma_8 = \frac{2}{3} \left(\kappa_W^2\!-\!c_{2V}\right)^2\! ,\quad
\Gamma_9 =-\frac{1}{3}\kappa_W \left(\kappa_W^2\!-\!c_{2V}\right) . 
\label{running2}
\end{align}
where only the first term in the expansion of $\Gamma_i$ in powers of $h/v$ is given, {\it i.e.}, $\Gamma_i (h=0)$. 
Note that $\Gamma_1=\Gamma_{3-9}=0$ and $\Gamma_2\neq  0$ for the SM values, 
$\kappa_W=c_{2V}=1$, as it should be.  



\begin{thebibliography}{90}

\bibitem{higgs}
  G.~Aad {\it et al.} [ATLAS Collaboration],
  ``Observation of a new particle in the search for the Standard Model Higgs boson with the ATLAS detector at the LHC,''
  Phys.\ Lett.\ B {\bf 716} (2012) 1
  [arXiv:1207.7214 [hep-ex]];
  S.~Chatrchyan {\it et al.} [CMS Collaboration],
  ``Observation of a New Boson at a Mass of 125 GeV with the CMS Experiment at the LHC,''
  Phys.\ Lett.\ B {\bf 716} (2012) 30
  [arXiv:1207.7235 [hep-ex]].

\bibitem{Buchalla:2016bse}
  G.~Buchalla, O.~Cat\`a, A.~Celis and C.~Krause,
  ``Standard Model Extended by a Heavy Singlet: Linear vs. Nonlinear EFT,''
  Nucl.\ Phys.\ B {\bf 917} (2017) 209
  [arXiv:1608.03564 [hep-ph]].

\bibitem{Pich:2018ltt}
  A.~Pich,
  ``Effective Field Theory with Nambu-Goldstone Modes,''
  arXiv:1804.05664 [hep-ph].

\bibitem{Pich:2015kwa}
  A.~Pich, I.~Rosell, J.~Santos and J.~J.~Sanz-Cillero,
  ``Low-energy signals of strongly-coupled electroweak symmetry-breaking scenarios,''
  Phys.\ Rev.\ D {\bf 93} (2016) no.5,  055041
  [arXiv:1510.03114 [hep-ph]].

  \bibitem{WSR}
  S.~Weinberg,
  ``Precise relations between the spectra of vector and axial vector mesons,''
  Phys.\ Rev.\ Lett.\  {\bf 18} (1967) 507.

 \bibitem{ST}
  A.~Pich, I.~Rosell and J.~J.~Sanz-Cillero,
  ``Viability of strongly-coupled scenarios with a light Higgs-like boson,''
  Phys.\ Rev.\ Lett.\  {\bf 110} (2013) 181801
  [arXiv:1212.6769 [hep-ph]];
%
  ``Oblique S and T Constraints on Electroweak Strongly-Coupled Models with a Light Higgs,''
  JHEP {\bf 1401} (2014) 157
  [arXiv:1310.3121 [hep-ph]].
  
 \bibitem{Peskin_Takeuchi}
  M.~E.~Peskin and T.~Takeuchi,
  ``A New constraint on a strongly interacting Higgs sector,''
  Phys.\ Rev.\ Lett.\  {\bf 65} (1990) 964;
  ``Estimation of oblique electroweak corrections,''
  Phys.\ Rev.\ D {\bf 46} (1992) 381.
  
  \bibitem{lagrangian_color}
  C.~Krause, A.~Pich, I.~Rosell, J.~Santos and J.~J.~Sanz-Cillero,
  ``Colorful Imprints of Heavy States in the Electroweak Effective Theory,''
  JHEP {\bf 1905} (2019) 092
  [arXiv:1810.10544 [hep-ph]].
  
    \bibitem{lagrangian}
  A.~Pich, I.~Rosell, J.~Santos and J.~J.~Sanz-Cillero,
  ``Fingerprints of heavy scales in electroweak effective Lagrangians,''
  JHEP {\bf 1704} (2017) 012
  [arXiv:1609.06659 [hep-ph]].
  
    \bibitem{Buchalla}
  G.~Buchalla, O.~Cat\`a and C.~Krause,
  ``Complete Electroweak Chiral Lagrangian with a Light Higgs at NLO,''
  Nucl.\ Phys.\ B {\bf 880} (2014) 552
   Erratum: [Nucl.\ Phys.\ B {\bf 913} (2016) 475]
  [arXiv:1307.5017 [hep-ph]];
  ``On the Power Counting in Effective Field Theories,''
  Phys.\ Lett.\ B {\bf 731} (2014) 80
  [arXiv:1312.5624 [hep-ph]].
  
  \bibitem{Weinberg}
  S.~Weinberg,
  ``Phenomenological Lagrangians,''
  Physica A {\bf 96} (1979) no.1-2,  327.
 
  \bibitem{Longhitano}
  A.~C.~Longhitano,
  ``Heavy Higgs Bosons in the Weinberg-Salam Model,''
  Phys.\ Rev.\ D {\bf 22} (1980) 1166;
  ``Low-Energy Impact of a Heavy Higgs Boson Sector,''
  Nucl.\ Phys.\ B {\bf 188} (1981) 118.

\bibitem{Herrero:1993nc}
  M.~J.~Herrero and E.~Ruiz Morales,
  ``The Electroweak chiral Lagrangian for the Standard Model with a heavy Higgs,''
  Nucl.\ Phys.\ B {\bf 418} (1994) 431
  [hep-ph/9308276].
  
\bibitem{Grinstein:2007iv}
  B.~Grinstein and M.~Trott,
  ``A Higgs-Higgs bound state due to new physics at a TeV,''
  Phys.\ Rev.\ D {\bf 76} (2007) 073002
  [arXiv:0704.1505 [hep-ph]].
  
  \bibitem{RChT}
  G.~Ecker, J.~Gasser, A.~Pich and E.~de Rafael,
  ``The Role of Resonances in Chiral Perturbation Theory,''
  Nucl.\ Phys.\ B {\bf 321} (1989) 311;
  G.~Ecker, J.~Gasser, H.~Leutwyler, A.~Pich and E.~de Rafael,
  ``Chiral Lagrangians for Massive Spin 1 Fields,''
  Phys.\ Lett.\ B {\bf 223} (1989) 425.

\bibitem{Filipuzzi:2012bv}
  A.~Filipuzzi, J.~Portol\'es and P.~Ruiz-Femen\'\i a,
  ``Zeros of the $W_L Z_L \rightarrow W_L Z_L$ Amplitude: Where Vector Resonances Stand,''
  JHEP {\bf 1208} (2012) 080
  [arXiv:1205.4682 [hep-ph]].

\bibitem{Alboteanu:2008my}
A.~Alboteanu, W.~Kilian and J.~Reuter,
``Resonances and Unitarity in Weak Boson Scattering at the LHC,''
JHEP \textbf{11} (2008), 010
[arXiv:0806.4145 [hep-ph]].

\bibitem{Toublan:1995bk}
D.~Toublan,
``Lowest tensor meson resonances contributions to the chiral perturbation theory low-energy coupling constants,''
Phys. Rev. D \textbf{53} (1996), 6602-6607
[arXiv:hep-ph/9509217 [hep-ph]].

\bibitem{Ecker:2007us}
G.~Ecker and C.~Zauner,
``Tensor meson exchange at low energies,''
Eur. Phys. J. C \textbf{52} (2007), 315-323
[arXiv:0705.0624 [hep-ph]].

\bibitem{Donoghue:1988ed}
J.~F.~Donoghue, C.~Ramirez and G.~Valencia,
``The Spectrum of QCD and Chiral Lagrangians of the Strong and Weak Interactions,''
Phys. Rev. D \textbf{39} (1989), 1947

\bibitem{Guo:2007ff}
Z.~Guo, J.~Sanz Cillero and H.~Zheng,
``Partial waves and large N(C) resonance sum rules,''
JHEP \textbf{06} (2007), 030
[arXiv:hep-ph/0701232 [hep-ph]].

\bibitem{Kilian:2015opv}
W.~Kilian, T.~Ohl, J.~Reuter and M.~Sekulla,
``Resonances at the LHC beyond the Higgs boson: The scalar/tensor case,''
Phys. Rev. D \textbf{93} (2016) no.3, 036004
[arXiv:1511.00022 [hep-ph]].

\bibitem{Frank:2012wh}
J.~Frank, M.~Rauch and D.~Zeppenfeld,
``Spin-2 resonances in vector-boson-fusion processes at next-to-leading order QCD,''
Phys. Rev. D \textbf{87} (2013) no.5, 055020
[arXiv:1211.3658 [hep-ph]].

\bibitem{Brass:2018hfw}
S.~Brass, C.~Fleper, W.~Kilian, J.~Reuter and M.~Sekulla,
``Transversal Modes and Higgs Bosons in Electroweak Vector-Boson Scattering at the LHC,''
Eur. Phys. J. C \textbf{78} (2018) no.11, 931
[arXiv:1807.02512 [hep-ph]].

\bibitem{Delgado:2017cls}
R.~Delgado, A.~Dobado, D.~Espriu, C.~Garcia-Garcia, M.~Herrero, X.~Marcano and J.~Sanz-Cillero,
``Production of vector resonances at the LHC via WZ-scattering: a unitarized EChL analysis,''
JHEP \textbf{11} (2017), 098
[arXiv:1707.04580 [hep-ph]].

\bibitem{Delgado:2019ucx}
R.~Delgado, C.~Garcia-Garcia and M.~Herrero,
``Dynamical vector resonances from the EChL in VBS at the LHC: the WW case,''
JHEP \textbf{11} (2019), 065
[arXiv:1907.11957 [hep-ph]].

\bibitem{Bernard:1975cd}
  C.~W.~Bernard, A.~Duncan, J.~LoSecco and S.~Weinberg,
  ``Exact Spectral Function Sum Rules,''
  Phys.\ Rev.\ D {\bf 12} (1975) 792.
  
  \bibitem{1stWSR}
  A.~Orgogozo and S.~Rychkov,
  ``Exploring T and S parameters in Vector Meson Dominance Models of Strong Electroweak Symmetry Breaking,''
  JHEP {\bf 1203} (2012) 046
  [arXiv:1111.3534 [hep-ph]];
  T.~Appelquist and F.~Sannino,
  ``The Physical spectrum of conformal SU(N) gauge theories,''
  Phys.\ Rev.\ D {\bf 59} (1999) 067702
  [hep-ph/9806409].
   
\bibitem{Aad:2019mbh}
  G.~Aad {\it et al.} [ATLAS Collaboration],
  ``Combined measurements of Higgs boson production and decay using up to $80$ fb$^{-1}$ of proton-proton collision data at $\sqrt{s}=$ 13 TeV collected with the ATLAS experiment,''
  Phys.\ Rev.\ D {\bf 101} (2020) no.1,  012002
  [arXiv:1909.02845 [hep-ex]].
  
\bibitem{Khachatryan:2014jba}
  V.~Khachatryan {\it et al.} [CMS Collaboration],
  ``Precise determination of the mass of the Higgs boson and tests of compatibility of its couplings with the standard model predictions using proton collisions at 7 and 8 $\,\text {TeV}$,''
  Eur.\ Phys.\ J.\ C {\bf 75} (2015) no.5,  212
  [arXiv:1412.8662 [hep-ex]].
  
\bibitem{deBlas:2018tjm}
  J.~de Blas, O.~Eberhardt and C.~Krause,
  ``Current and Future Constraints on Higgs Couplings in the Nonlinear Effective Theory,''
  JHEP {\bf 1807} (2018) 048
  [arXiv:1803.00939 [hep-ph]].
  
\bibitem{ATLAS:2019dgh}
  The ATLAS collaboration [ATLAS Collaboration],
  ``Search for the $HH \to b \bar{b} b \bar{b}$ process via vector boson fusion production using proton-proton collisions at $\sqrt{s}$ = 13 TeV with the ATLAS detector,''
  ATLAS-CONF-2019-030.
 
\bibitem{Delgado:2014jda}
  R.~L.~Delgado, A.~Dobado, M.~J.~Herrero and J.~J.~Sanz-Cillero,
  ``One-loop $\gamma\gamma \to$ W$_{L}^{+}$ W$_{L}^{-}$ and $\gamma\gamma \to$ Z$_{L}$ Z$_{L}$ from the Electroweak Chiral Lagrangian with a light Higgs-like scalar,''
  JHEP {\bf 1407} (2014) 149
  [arXiv:1404.2866 [hep-ph]].
  
\bibitem{Tanabashi:2018oca}
  M.~Tanabashi {\it et al.} [Particle Data Group],
  ``Review of Particle Physics,''
  Phys.\ Rev.\ D {\bf 98} (2018) no.3,  030001.

\bibitem{Falkowski:2015jaa}
  A.~Falkowski, M.~Gonz\'alez-Alonso, A.~Greljo and D.~Marzocca,
  ``Global constraints on anomalous triple gauge couplings in effective field theory approach,''
  Phys.\ Rev.\ Lett.\  {\bf 116} (2016) no.1,  011801
  [arXiv:1508.00581 [hep-ph]].
  
\bibitem{Almeida:2018cld}
  E.~da Silva Almeida, A.~Alves, N.~Rosa Agostinho, O.~J.~P.~Éboli and M.~C.~Gonz\'alez-Garc\'\i a,
  ``Electroweak Sector Under Scrutiny: A Combined Analysis of LHC and Electroweak Precision Data,''
  Phys.\ Rev.\ D {\bf 99} (2019) no.3,  033001
  [arXiv:1812.01009 [hep-ph]].
  
\bibitem{Sirunyan:2019der}
  A.~M.~Sirunyan {\it et al.} [CMS Collaboration],
  ``Search for anomalous electroweak production of vector boson pairs in association with two jets in proton-proton collisions at 13 TeV,''
  Phys.\ Lett.\ B {\bf 798} (2019) 134985
  [arXiv:1905.07445 [hep-ex]].

\bibitem{Rauch:2016pai}
  M.~Rauch,
  ``Vector-Boson Fusion and Vector-Boson Scattering,''
  arXiv:1610.08420 [hep-ph].
  
\bibitem{claudia}
  C.~Garc\'\i a-Garc\'\i a,
  arXiv:2001.08965 [hep-ph].

\bibitem{Garcia-Garcia:2019oig}
  C.~Garc\'\i a-Garc\'\i a, M.~Herrero and R.~A.~Morales,
  ``Unitarization effects in EFT predictions of WZ scattering at the LHC,''
  Phys.\ Rev.\ D {\bf 100} (2019) no.9,  096003
  [arXiv:1907.06668 [hep-ph]].

\bibitem{Fabbrichesi:2015hsa}
  M.~Fabbrichesi, M.~Pinamonti, A.~Tonero and A.~Urbano,
  ``Vector boson scattering at the LHC: A study of the WW $\to$ WW channels with the Warsaw cut,''
  Phys.\ Rev.\ D {\bf 93} (2016) no.1,  015004
  [arXiv:1509.06378 [hep-ph]].

\bibitem{Sirunyan:2020gyx}
A.~M.~Sirunyan \textit{et al.} [CMS],
``Measurements of production cross sections of WZ and same-sign WW boson pairs in association with two jets in proton-proton collisions at $\sqrt{s}=$ 13 TeV,''
[arXiv:2005.01173 [hep-ex]].

\bibitem{Perez:2018kav}
G.~Perez, M.~Sekulla and D.~Zeppenfeld,
``Anomalous quartic gauge couplings and unitarization for the vector boson scattering process $pp\rightarrow W^+W^+jjX\rightarrow \ell ^+\nu _\ell \ell ^+\nu _\ell jjX$,''
Eur. Phys. J. C \textbf{78} (2018) no.9, 759
[arXiv:1807.02707 [hep-ph]].

\bibitem{Szleper:2014xxa}
M.~Szleper,
``The Higgs boson and the physics of $WW$ scattering before and after Higgs discovery,''
[arXiv:1412.8367 [hep-ph]].

\bibitem{Kalinowski:2018oxd}
J.~Kalinowski, P.~Kozów, S.~Pokorski, J.~Rosiek, M.~Szleper and S.~Tkaczyk,
``Same-sign WW scattering at the LHC: can we discover BSM effects before discovering new states?,''
Eur. Phys. J. C \textbf{78} (2018) no.5, 403
[arXiv:1802.02366 [hep-ph]].

\bibitem{Kozow:2019txg}
P.~Kozów, L.~Merlo, S.~Pokorski and M.~Szleper,
``Same-sign WW Scattering in the HEFT: Discoverability vs. EFT Validity,''
JHEP \textbf{07} (2019), 021
[arXiv:1905.03354 [hep-ph]].

\bibitem{Guo:2015isa}
  F.~K.~Guo, P.~Ruiz-Femen\'\i a and J.~J.~Sanz-Cillero,
  ``One loop renormalization of the electroweak chiral Lagrangian with a light Higgs boson,''
  Phys.\ Rev.\ D {\bf 92} (2015) 074005
  [arXiv:1506.04204 [hep-ph]].

\bibitem{Sikivie:1980hm}
  P.~Sikivie, L.~Susskind, M.~B.~Voloshin and V.~I.~Zakharov,
  ``Isospin Breaking in Technicolor Models,''
  Nucl.\ Phys.\ B {\bf 173} (1980) 189.
 
 \bibitem{CCWZ} 
  S.~R.~Coleman, J.~Wess and B.~Zumino,
  ``Structure of phenomenological Lagrangians. 1.,''
  Phys.\ Rev.\  {\bf 177} (1969) 2239;
  C.~G.~Callan, Jr., S.~R.~Coleman, J.~Wess and B.~Zumino,
  ``Structure of phenomenological Lagrangians. 2.,''
  Phys.\ Rev.\  {\bf 177} (1969) 2247.
 
\bibitem{Pich:2012jv}
  A.~Pich, I.~Rosell and J.~J.~Sanz-Cillero,
  ``One-Loop Calculation of the Oblique S Parameter in Higgsless Electroweak Models,''
  JHEP {\bf 1208} (2012) 106
  [arXiv:1206.3454 [hep-ph]].





\end{thebibliography}
\end{document}